\documentclass[twocolumn,dvipsnames]{aastex631}

\usepackage{amsmath}
\usepackage{booktabs}
\usepackage{array}

\newcommand{\CDF}{CDF50}
\newcommand{\Peaks}{PDFPeak}

\begin{document}

\title{Astrometric Redshifts of Supernovae}

\newcommand{\jl}[1]{\textcolor{ForestGreen}{[{\bf JL}: #1]}}
\newcommand{\mscomment}[1]{\textcolor{red}{[{\bf MS}: #1]}}

\author[0000-0001-6633-9793]{Jaemyoung (Jason) Lee}
\thanks{astjason@sas.upenn.edu}
\affiliation{Department of Physics and Astronomy, University of Pennsylvania, Philadelphia, PA 19104, USA}
\author[0000-0003-2764-7093]{Masao Sako}
\affiliation{Department of Physics and Astronomy, University of Pennsylvania, Philadelphia, PA 19104, USA}
\author[0000-0003-3221-0419]{Richard Kessler}
\affiliation{Department of Astronomy and Astrophysics, University of Chicago, Chicago, IL 60637, USA}
\author[0000-0002-8676-1622]{Alex I. Malz}
\affiliation{McWilliams Center for Cosmology, Department of Physics, Carnegie Mellon University, Pittsburgh, PA 15213, USA}
\author{The LSST Dark Energy Science Collaboration}



\begin{abstract}

Differential Chromatic Refraction (DCR) is caused by the wavelength dependence of our atmosphere's refractive index, which shifts the apparent positions of stars and galaxies and distorts their shapes depending on their spectral energy distributions (SEDs).~While this effect is typically mitigated and corrected for in imaging observations, we investigate how DCR can instead be used to our advantage to infer the redshifts of supernovae from multi-band, time-series imaging data.~We simulate Type Ia supernovae (SNe Ia) in the proposed Vera C. Rubin Observatory Legacy Survey of Space and Time (LSST) Deep Drilling Field (DDF),
and evaluate astrometric redshifts.~We find that the redshift accuracy improves dramatically with the statistical quality of the astrometric measurements as well as with the accuracy of the astrometric solution. For a conservative choice of a 5-mas systematic uncertainty floor, we find that our redshift estimation is accurate at $z < 0.6$.  We then combine our astrometric redshifts with both host-galaxy photometric redshifts and supernovae photometric (light-curve) redshifts and show that this considerably improves the overall redshift estimates.~These astrometric redshifts will be valuable especially since Rubin will discover a vast number of supernovae for which we will not be able to obtain spectroscopic redshifts.

\end{abstract}



\section{Introduction} \label{sec:intro}

Observations of Type Ia supernovae (SNe) resulted in the unexpected but groundbreaking discovery of the accelerating universe \citep{riess1998,perlmutter1999}.~While this discovery with only tens of SNe Ia is an extraordinary feat, subsequent efforts have allowed for increasingly precise constraints on the dark energy equation of state parameter $w$ and the matter density $\Omega_m$.~Indeed, recent measurements from Pantheon+ \citep{brout2022pantheon+} and the Dark Energy Survey (DES) \citep{DES-SN5YR_Key2024} now give roughly $0.15$ uncertainty on the equation of state parameter $w_0$ for a flat $w_0$CDM model from over 1000 SNe each. Combining supernovae measurements with Cosmic Microwave Background (CMB) and Baryon Acoustic Oscillations (BAO) measurements, which are largely independent from SNe Ia measurements, significantly improves the cosmological constraints. 

Accurate redshifts are crucial for SN Ia cosmology. Up to this date, SN redshifts have relied on spectroscopic redshifts, either of the SN itself or from the host-galaxy \citep{howell2005snls_gemini,sako2018sdssII,lidman2020ozdes_spectroscopy,smith2020DES_characteristics}. However, the Vera C. Rubin Observatory Legacy Survey of Space and Time \citep[LSST,][]{ivezic2019lsst} will observe orders of magnitude more SNe than all of the SNe detected so far, making it no longer viable to perform spectroscopic follow-up for all of the detected SNe and their hosts. 

Consequently, there have been numerous efforts to utilize SN photometry to accurately measure redshifts. Some of these efforts include using the peak flux in the observed bands as inputs for an empirical, analytic redshift estimator \citep{wang2006model_independent_photoz,wang2007_survey_requirements,wang2015LSST_photoZ}, adding the redshift as an additional light-curve fitting parameter, also known as \verb|LCFIT+z| \citep{kessler2010photo_z,PD_photoz}, and using a machine learning approach to learn the redshifts from light-curves \citep{qu2023photo_z_ML}.~Many studies have sought to improve SN photometric redshift measurements by adding a host-galaxy photometric redshift  (photo-$z$) prior, including three of the aforementioned works, as well as \citet{mitra2023LSST_photoz_with_hostgal}, where the resulting improvement in cosmological constraints were explored extensively.

\begin{figure}
    \centering
    \includegraphics[width=0.48\textwidth]{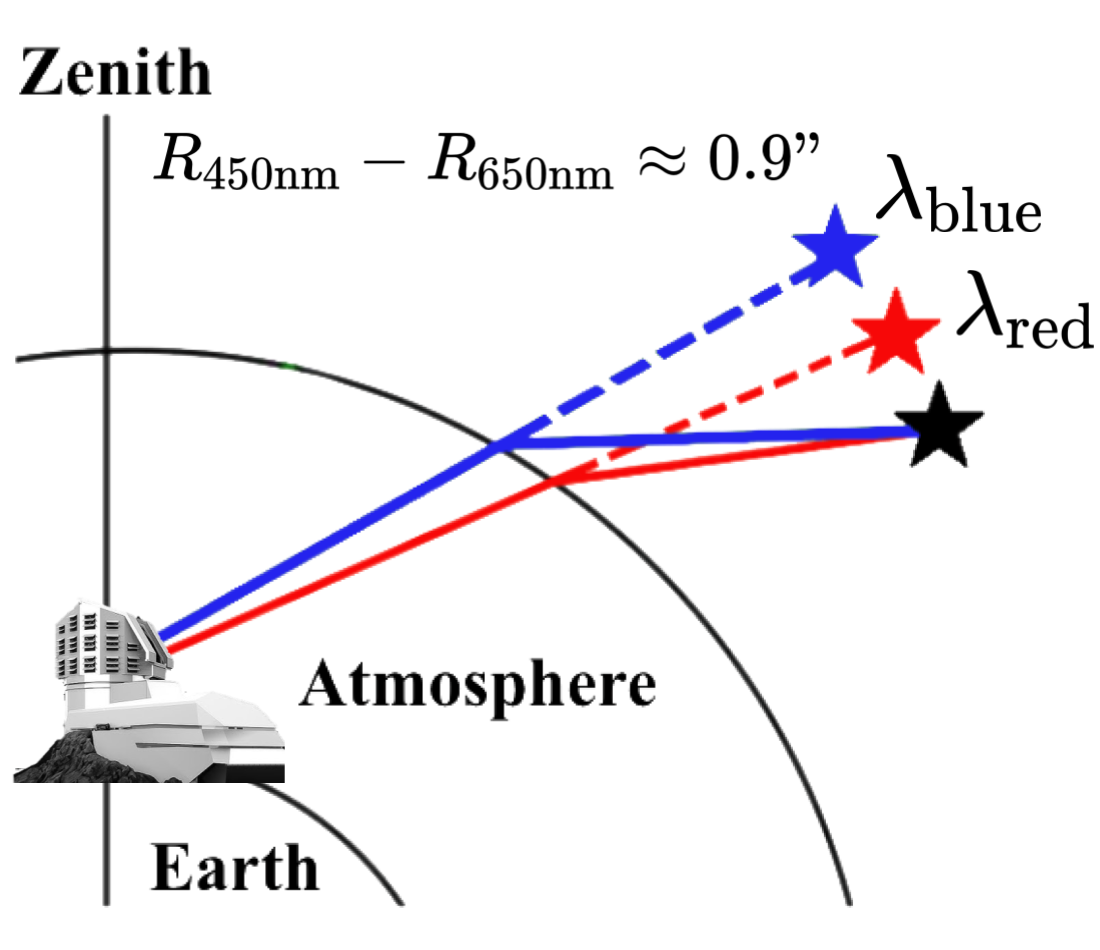}
    \caption{A schematic diagram showing how DCR affects light in the Earth’s atmosphere; shorter wavelengths are refracted more by the atmosphere, which means that to an observer on Earth, a blue star appears to be higher up in the sky than a red star located at the same position in space. $R_{450\rm{nm}}$ and $R_{650\rm{nm}}$ denote the amount of refraction by our atmosphere at an air mass (AM) of 1.4 for 450 nm and 650 nm light respectively.}
    \label{fig:DCR_diagram}
\end{figure}

While the impact of photometric redshift uncertainties on cosmology may not be severe \citep{dai2018LSST_photoz,chen2022redmagic,mitra2023LSST_photoz_with_hostgal}, better redshift measurements will lead to more constraining power in the cosmological parameter estimation by reducing the uncertainties in the redshifts used for cosmology.~Here, we introduce another method to obtain redshift measurements using imaging data.~This method utilizes the fact that sources at different redshifts (and hence different colors) are refracted by different amounts in our atmosphere.~As illustrated in Figure \ref{fig:DCR_diagram}, shorter wavelength light is refracted more in our atmosphere than longer wavelength light. This effect, known as Differential Chromatic Refraction \citep[DCR,][]{filippenko1982DCR}, is a result of the refractive index of our atmosphere being wavelength dependent. 

Normally DCR is an effect cosmologists want to mitigate for precision cosmology in both weak lensing \citep{plazas2012lambda_dependent_WL,meyers2015_DCR,carlsten2018_DCR} and SN Ia cosmology \citep*{DES5YR-DCR}. However, \citet{kaczmarczik2009astro_z} found that treating the observed DCR shifts in the Sloan Digital Sky Survey (SDSS) \textit{u} and \textit{g} bands as additional colors in photometric redshifts improves quasar photometric redshifts, with the fraction of objects within $\Delta z = \pm 0.1$ of the spectroscopic redshifts increasing by 9\% and catastrophic outliers reduced.~As such, astrometric redshifts have the potential to strengthen photo-$z$ constraints when combined with photo-$z$'s.~They also mention the possibility of applying these \textit{astrometric redshifts} (astro-$z$'s) to other sources with distinct spectral features such as Type Ia and Type II supernovae. 

In this work, we develop the methodology to obtain astrometric redshifts for Type Ia supernovae. For this development, we added DCR effects to the \verb|SNANA|\footnote{https://github.com/RickKessler/SNANA} \citep[SuperNova ANAlysis,][]{kessler2009snana} simulation, and analyzed simulated SN Ia samples corresponding to the proposed Deep Drilling Fields (DDF) for LSST. We show astro-$z$ constraints from only the DCR effect, utilizing multi-band, multi-epoch measurements, and we show how combining them with both host and SN photo-$z$'s improves the overall redshift estimates.


The content of this work is as follows. In Section \ref{sec:DCR}, we compute the amount of DCR shift that occurs depending on the air mass, source flux and filter response functions and describe how SNe at different redshifts are affected. Next, we describe the simulated LSST dataset in Section \ref{sec:data_sims}.~Then, we describe our methods to obtain redshift estimates from the simulated astrometry for each SN observation in Section \ref{sec:methods}.~In Section \ref{sec:results}, we show our results for LSST-like simulations in hypothetical (ideal) and realistic cases, as well as the improvement when combined with both host-galaxy and SNe photometric redshifts, and extensively discuss our results. 
Lastly, we end with a discussion and conclusion in Section \ref{sec:conclusion}. 

\begin{figure}
    \centering
    \includegraphics[width=0.48\textwidth]{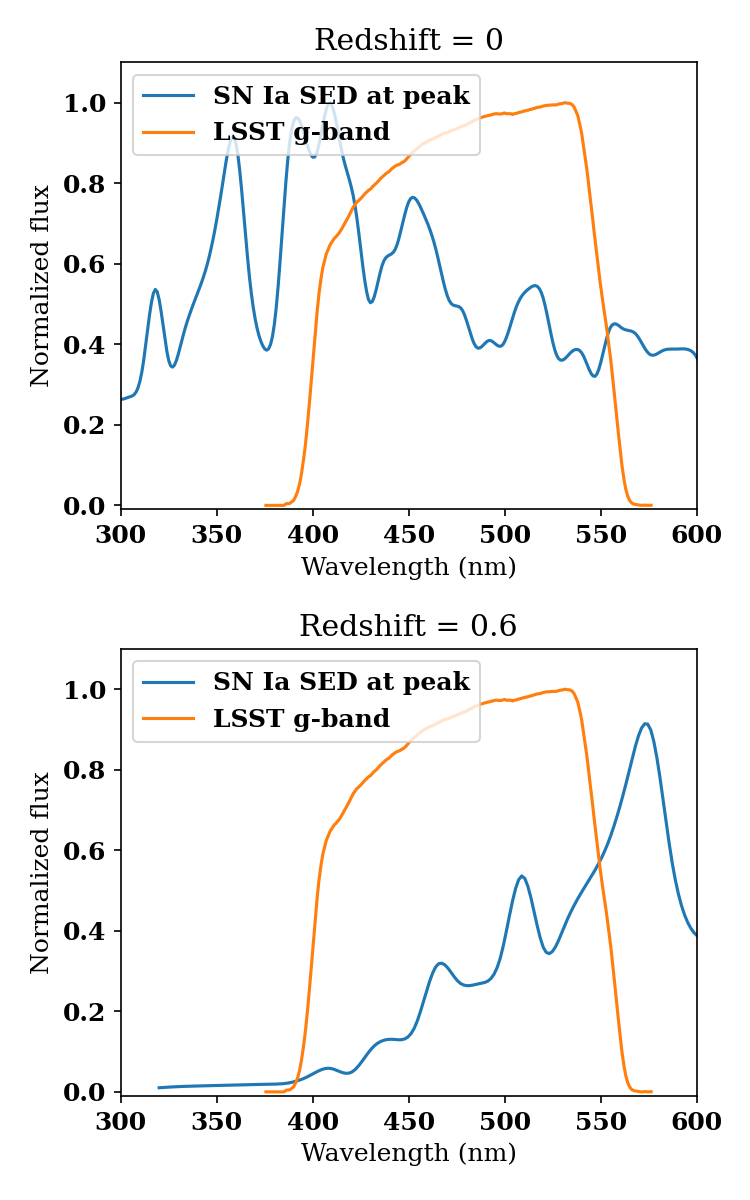}
    \caption{Example SN Ia SED at peak epoch overplotted on top of the LSST \textit{g}-band at $z = 0$ (top) and $z = 0.6$ (bottom). As shown in Equation \ref{eq:DCR_alt_shift}, the resulting DCR shifts at the two redshifts are different since the overlap between the filter and the SED changes significantly.~Because the product of the filter function and the SED, $F(\lambda)S(\lambda)$, is smaller for the $z = 0.6$ SN, the (absolute) DCR shift is also smaller, as shown in Figure~\ref{fig:ALT_shift_redshift}.}
    \label{fig:SED_and_LSST_g}
\end{figure}

\section{Differential Chromatic Refraction} \label{sec:DCR}

DCR is due to the dependency of our atmosphere's refractive index on the source wavelength. The centroid shift caused by DCR can be calculated as in \citet{plazas2012lambda_dependent_WL}: 
\begin{equation}
    \bar{R} = \frac{\int d\lambda R(\lambda, z_a)F(\lambda) \lambda S(\lambda)}{\int d\lambda F(\lambda)  \lambda S(\lambda)}
    \label{eq:DCR_alt_shift}
\end{equation} 
\noindent where $\bar{R}$ is the altitude shift of the source towards the zenith, $R(\lambda, z_a)$ is the shift of each photon towards the zenith, $z_a$ is the zenith angle, $F(\lambda)$ is the filter response function, and $S(\lambda)$ is the source spectral energy distribution (SED).\footnote{Note that the factor of $\lambda$ is necessary in the integrands of both the numerator and denominator if $S(\lambda)$ is taken to be the spectral \textit{energy} distribution. This takes into account that \textit{photon-counting} detectors are used for our observations.} $R(\lambda, z_a)$ can be calculated using: 
\begin{equation}
    R(\lambda, z_a) \approx R_0 \tan(z_a), \quad R_0 = \frac{n^2(\lambda)-1}{2n^2(\lambda)}
\end{equation} \label{eq:DCR_photon_shift}

\noindent with the air mass (AM), or the amount of air along the line of sight, being $\mathrm{AM} = \sec z_a$ for $z_a < 80^{\circ}$. For an explicit expression of $n(\lambda)$, see \citet{filippenko1982DCR}.~As apparent in Equation \ref{eq:DCR_alt_shift}, the DCR altitude shift depends on the filter transmission function.~It also depends on the source SED, which depends on both redshift and the epoch of the SN. The dependency on $R(\lambda, z_a)$ means that the shift becomes larger at higher AM. 

\begin{figure}
    \centering
    \includegraphics[width=0.48\textwidth]{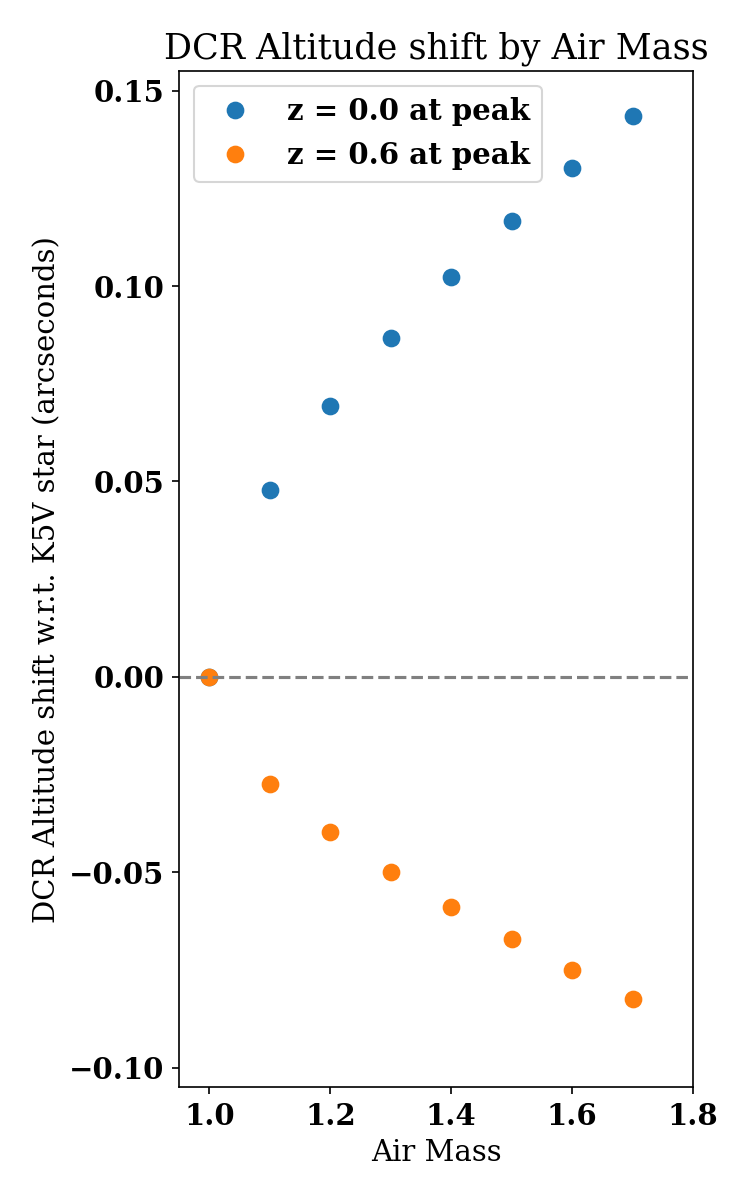}
    \caption{The resulting \textit{g}-band DCR altitude shifts with respect to a K5V reference star for a $z = 0$ SN (blue) and $z = 0.6$ SN (orange) at different air masses (AMs), but both at peak epoch.~As we expect, the shifts at the two redshifts are different and we see more shift at higher AM. Comparison with Figure \ref{fig:SED_and_LSST_g} implies that the $z = 0$ SN is ``bluer'' (in the \textit{g}-band at least) than the reference star while the $z = 0.6$ SN is ``redder.''}
    \label{fig:ALT_shift_redshift}
\end{figure}

Figure \ref{fig:SED_and_LSST_g} shows that as the SN redshift increases, the overlap of its SED with the filter function changes, so the DCR altitude shift changes with redshift. Although the amount of refraction that occurs due to DCR is usually between 43$\arcsec$ to 46$\arcsec$ for a given source viewed through the LSST \textit{ugriz} bands at an AM of 1.4, the observed shifts are much smaller because DCR also affects
the reference stars that are used to determine the astrometric solution.~Hence, we measure the shift with respect to the average color star in a given CCD, which we take in this work to be a K5V star, adopted from the DES-SN5YR analysis \citep*{DES5YR-DCR}. For LSST, we do not know beforehand the spectrum of the average color star, but it is straightforward to use another spectral type as the reference star.~Since the reference stellar type only determines the amplitudes of the measured DCR shifts, our subsequent results are not sensitive to the choice of reference stellar type unless its spectra is significantly different from the one chosen here.~Figure \ref{fig:ALT_shift_redshift} shows the \textit{g}-band altitude shift vs. AM for a SN (with the light-curve stretch and color being 0) at rest-frame epoch $T_{\rm{rest}} = 0$ for $z = 0$ and $z = 0.6$ where $T_{\rm{rest}}$ is the number of days from the time of peak brightness in the \textit{g}-band.~Compared to the average color star, the $z = 0$ SN is DCR shifted upwards (towards zenith) while the $z = 0.6$ SN is shifted downwards (away from zenith) at $\rm{AM} > 1.0$. We also show the expected increase in the DCR shifts at higher AM. 

\section{Dataset and Simulations} \label{sec:data_sims}

We perform our analysis on a simulated data sample generated assuming LSST survey characteristics. LSST is a ground-based survey using the 8.4 m Simonyi Survey Telescope in Chile with expected first light in 2025. Over the period of a decade, LSST will observe 18,000 $\rm{deg}^2$ of the southern sky with its 3.2-gigapixel camera and a very wide 9.6 $\rm{deg}^2$ field of view in six filters \textit{ugrizY} covering $320$ to $1050$ nm. We only use the Deep Drilling Fields (DDF) subsample \citep{scolnic2018lsst_DDF} obtained by coadding exposures from each of the bands taken within the same night. 


We simulate LSST-like SNe Ia observations in the \textit{ugrizY} bands using \verb|SNANA|, which uses LSST-like SN Ia observing characteristics following the Extended LSST Astronomical Time-series Classification Challenge (ELAsTiCC) dataset \citep{ELAsTiCC2023}, which is a sequel to the earlier PLAsTiCC dataset \citep{allam2018PLAsTiCC,kessler2019PLAsTiCC}.~The simulation includes realistic SN observations with a source model that generates SEDs for each epoch, galactic extinction that impact the SED, and a noise model that accounts for the PSF, sky noise and zeropoint as highlighted in \citet{kessler2019SNANA}. 


For our SED model in the simulation and analysis, 
we use the SALT3 model \citep{kenworthy2021salt3} with wavelengths extended to the near-infrared (coverage up to to 2,000 nm) following \citep{pierel2022salt3}.~The SALT2 \citep{guy2007salt2} and SALT3 models have a known artifact of producing negative SED fluxes in the ultraviolet region for rest-frame epochs $T_{\mathrm{rest}} < -15$ days.~Better quality data in the future will hopefully fix this artifact, but here we simply force negative fluxes to be zero.~We note that observations at $T_{\mathrm{rest}} < -15$ days are rare occurrences; less than 1\% of our simulated observations are at $T_{\mathrm{rest}} < -15$ days.~Host-galaxy properties are modeled as described in \citet{lokken2023simulated}. 




\begin{figure}
    \centering
    \includegraphics[width=0.48\textwidth]{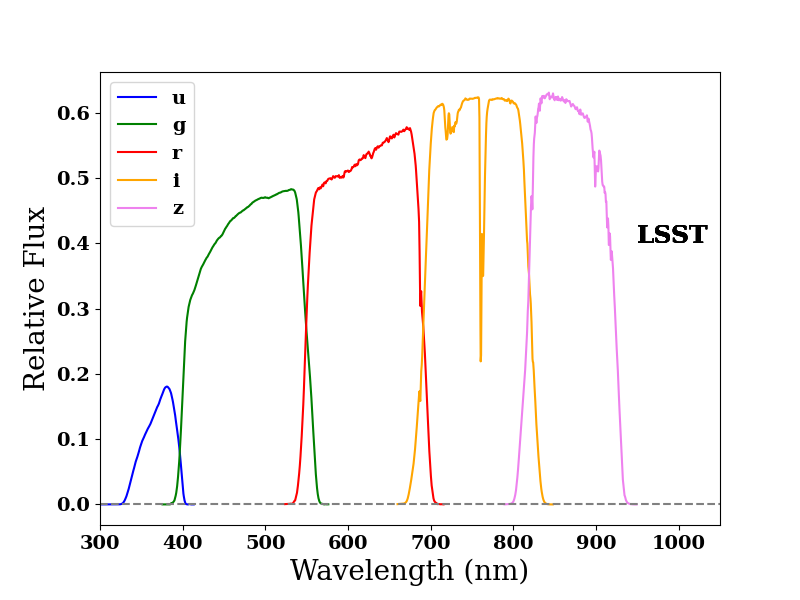}
    \caption{LSST \textit{ugriz} filters (from left to right) according to their relative transmission and wavelength range.~We do not show the \textit{Y} band as it is not used for the DCR calculations.~Multiple detections with different bands are crucial for accurate  astrometric redshifts.}
    \label{fig:LSST_filters}
\end{figure}

\begin{figure*}
    \centering
    \includegraphics[width=0.96\textwidth]{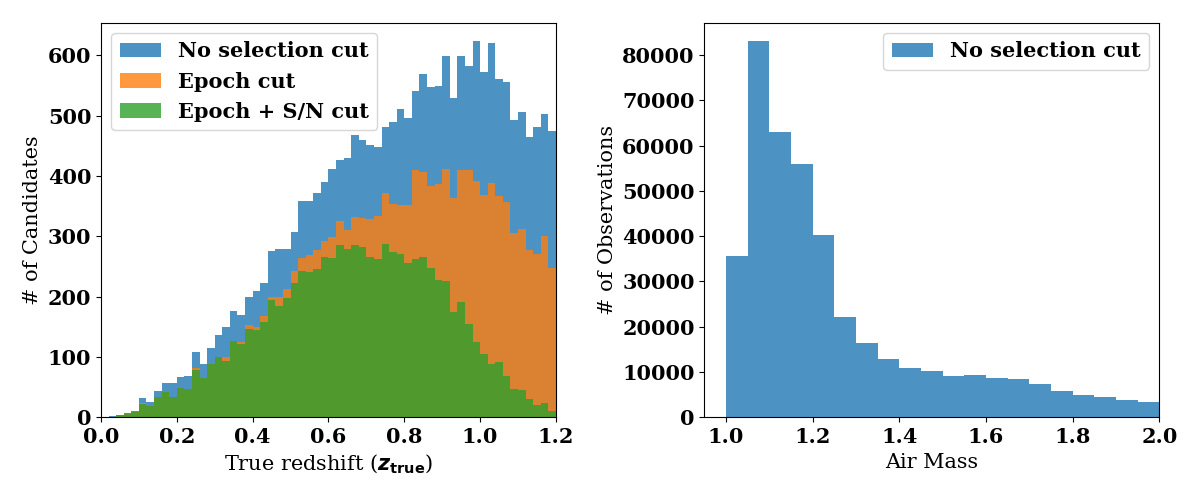}
    \caption{Redshift and air mass (AM) distributions for the LSST DDF simulations in SNANA. The peak redshift is around $z = 0.9$ for No selection cut and Epoch cut, but around $0.6 < z < 0.9$ for the Epoch + S/N cut, while the AM peaks between 1.05 and 1.10 for all three selection cuts. Higher AM observations lead to better astro-$z$ estimates.}
    \label{fig:redshift_and_AM_LSST}
\end{figure*}

Some survey characteristics particularly important for this work include the filter response functions\footnote{\url{https://github.com/lsst/throughputs/tree/1.9/baseline}} shown in Figure \ref{fig:LSST_filters}, and the simulated air mass as well as  redshift distributions, shown in Figure \ref{fig:redshift_and_AM_LSST}. Because the DCR shifts are typically the largest in the \textit{u} and \textit{g}-bands, observations in those bands will be especially helpful in measuring SN redshifts. The redshift distribution peaks around $z = 0.9$ without selection requirements (cuts; see end of Section \ref{sec:data_sims}), but at $0.6 < z < 0.9$ with cuts. Most of the observations are at $1.05 < \rm{AM} < 1.25$, regardless of the selection cut, meaning that the vast majority of the SNe observations will result in some sort of shift away from the reference star.~Roughly 23\% of the observations are at high 
AM ($> 1.4$) where we see the most extreme shifts. 


To simulate realistic astrometry, we added new functionalities in the \verb|SNANA| simulations.~We simulate true DCR shifts in the Right Ascension (RA, $\alpha$) and Declination (DEC, $\delta$) directions by first calculating the DCR altitude shifts for the SN ($\bar{R}_{\rm{ALT, SN}}$) and K5V star ($\bar{R}_{\rm{ALT, K5V}}$) by performing the integral given in Equation \ref{eq:DCR_alt_shift} then adding $\Delta_{\rm{DCR}} \alpha$ and $\Delta_{\rm{DCR}} \delta$ to the reference star (K5V) RA and DEC after DCR where 
\begin{equation}
\begin{split}
    \Delta_{\rm{DCR}} \alpha & = (\bar{R}_{\rm{ALT, SN}} - \bar{R}_{\rm{ALT, K5V}}) \frac{\sin q}{\cos \delta} \\
\Delta_{\rm{DCR}} \delta & = (\bar{R}_{\rm{ALT, SN}} - \bar{R}_{\rm{ALT, K5V}}) \cos{q}
\end{split}
\end{equation} \label{eq:RA_DEC_DCR_shift}
\noindent and $q$ is the parallactic angle.~While Equation \ref{eq:RA_DEC_DCR_shift} is an approximation, the true shift calculated using spherical trigonometry results in a very similar answer.~This is because the shifts considered are smaller than $1\arcsec$.~We show in Appendix \ref{Appendix:DCR_calc} a diagram of the celestial sphere for an observer in the Southern Hemisphere as well as detailed calculations of the shifts. 

Next, because real observations include Poisson noise, the SN centroid accuracy depends on the S/N of the SN measurement.~Thus, we add random Gaussian smearing as a function of ($\rm{FWHM}_{\rm{PSF}}$/[S/N]) for our statistical uncertainties.~To parametrize the angular resolution, we use the GalSIM\footnote{\url{https://galsim-developers.github.io/GalSim/\_build/html/index.html}} \citep{rowe2015galsim} package to simulate images with \textit{only} Poisson noise, calculate the standard deviation of RA and DEC shifts depending on the $\rm{FWHM}_{\rm{PSF}}$ and S/N at multiple AMs, and fit a 2nd-order polynomial of ($\rm{FWHM}_{\rm{PSF}}$($\arcsec$)/[S/N]). Ideally, we would use GalSIM to simulate each SN observation including Poisson noise, DCR effects, as well as the host-galaxy and measure the centroid shift for each observation.~However, this is computationally expensive, especially if we were to use the full scene-modeling pipeline that forward models the SN Ia and host-galaxy as done for typical SN Ia surveys \citep{astier2006snls_SMP,brout2019SMP,bernardinelli2023photometry}, which can take tens of hours per candidate.~To reduce the computation time, we chose to simulate the random shift arising from Poisson noise and DCR instead of simulating each image in this initial investigation.~The quadrature sum of our fitted polynomial function for the standard deviation and a systematic floor gives the total uncertainties, $\sigma_{\mathrm{tot}}^2 = \sigma_{\mathrm{syst}}^2 + \sigma_{\mathrm{stat}}^2$, where
\begin{equation} \label{eq:sigma_stat}
   \sigma_{\mathrm{stat}} = 0.66 \left(\frac{\rm{FWHM}_{\rm{PSF}}(\arcsec)}{\rm{[S/N]}}\right) + \frac{0.1}{1 \arcsec} \left(\frac{\rm{FWHM}_{\rm{PSF}}(\arcsec)}{\rm{[S/N]}}\right) ^2
\end{equation} 

\noindent with $\sigma_{\rm{syst}} = 5$ mas.~The simulated random RA and DEC shifts due to the total uncertainties are: 

\begin{equation}
    \Delta_{\sigma, \rm{tot}} \alpha = \sigma_{\rm{tot}}  \frac{\mathcal{N}_{\alpha}(0, 1)}{\cos{\delta}}, \quad \Delta_{\sigma, \rm{tot}} \delta = \sigma_{\rm{tot}}   \mathcal{N}_{\delta}(0, 1)
\end{equation} \label{eq:RA_DEC_sigma_tot_shift}

\noindent where $\mathcal{N}_{\alpha}(0, 1)$ and $\mathcal{N}_{\delta}(0, 1)$ are random Gaussian deviates each with a mean of zero and standard deviation of 1. Combining everything, we have: 
\begin{equation}
    \alpha_{\rm{obs}} = \alpha_{\rm{true}} + \Delta_{\rm{DCR}} \alpha + \Delta_{\sigma, \rm{tot}} \alpha 
\end{equation} \label{eq:RA_obs_tot}

\noindent and likewise for $\delta_{\rm{obs}}$.~When the S/N is too low, $\sigma_{\rm{stat}}$ becomes larger than the largest DCR shifts we observe, or around $0.1\arcsec$ to $0.2\arcsec$, which typically occur at high AM observations in the \textit{ug}-bands at $z < 0.1$.~Thus, we only use DCR shifts when the S/N is larger than 3 ($\sigma_{\rm{stat}} \approx 0.23\arcsec$ for a $1\arcsec$ PSF when S/N = 3).~It is possible in theory to use all observations including when $\mathrm{S/N} < 3$, but this is likely not feasible for real observations due to the difficulty of accurately fitting for centroids at such low S/N. Because Poisson noise causes the centroids to shift in random directions, using only $\mathrm{S/N} > 3$ SNe for our analysis will not result in a bias.

To compare the simulated DCR measurement and the model DCR shift on a given epoch, we subtract the peak MJD of the SN from its observed date and take this value to be the observed epoch $T_{\rm{obs}}$.~To estimate peak MJDs, we use the `Fmax-clump' method as described in the \verb|SNANA| manual, where peak MJD is defined to be the epoch of maximum flux in the sliding 50-day window containing the most detections.~This method is robust because there is no fitting, and it ignores pathological fluxes that are far away
from peak brightness.~Next, we use the light-curve fitting module \verb|LCFIT+z| \citep{kessler2010photo_z} with a weak cosmology prior as described in Appendix \ref{Appendix:cosmology_prior}, only for the purpose of obtaining light-curve fit peak MJD values that are input to the astro-$z$ pipeline. Since we use the very high-cadence DDF only, the `Fmax-clump' method is adequate.~However, for surveys with poorer cadence (e.g., WFD), light-curve fit may be needed to get a better estimate of peak MJDs. Because the light-curve fitting is not always successful, this process leaves us with 13,735 out of 20,000 candidates, or 7,804 out of the 8,588 candidates that pass the selection cuts as described later in this section.~The primary reason for unsuccessful light-curve fitting is not having observations pre- and post-peak epoch, while having low S/N is another reason.~Thus, we combine the astro-$z$ measurements obtained using \verb|LCFIT+z| peak MJDs (when available) with `Fmax-clump' peak MJDs (when \verb|LCFIT+z| peak MJDs are not available) when showing our default results in Section \ref{sec:realistic}. We also show some performance metrics i) with only `Fmax-clump' peak MJDs and ii) assuming we know precisely the peak MJDs in Table \ref{tab:statistics_realistic_by_selection_cut}. 


Our simulation does not account for other wavelength dependent effects in the telescope optics and detectors affecting SN centroid measurements. Some of these effects were investigated in detail for DECam in \citet{bernstein2017astrometric,bernstein2018photometric}. 
The largest optics effect comes from refraction in the lenses, shifting sources in different bands by different amounts particularly near the boundaries of the focal plane.~Such color-dependent radial displacements (also known as lateral color) are shown to be around $0.050\arcsec$/mag and $0.005\arcsec$/mag in the \textit{g} and \textit{r} bands in $g-i$ magnitudes respectively near the edges of the focal plane for DES. While \citet*{DES5YR-DCR} mentions that these shifts are at worst, similar to the DCR positional effects for DES, it is plausible to expect that at least some of these effects will be corrected for in LSST. For DECam, the uncertainty in the astrometric solution due to telescope optics such as unmodeled stray electric fields in the detectors as well as focal plane shifts between camera thermal cycles is about 3-6 mas on the focal plane \citep{bernstein2017astrometric}, so our assumption of a $\sigma_{\mathrm{syst}} = 5$ mas systematic floor is plausible.~Therefore, the uncertainties we model in this study will account for the bulk of the uncertainties we will face in real observations. 


Because we advocate the combination of photo-$z$'s with our astro-$z$'s, we apply selection cuts typically applied to photo-$z$'s to our simulated samples when showing results for realistic simulations.~Following \citet{kessler2010photo_z}, we first require that there are observations before $T_{\rm{rest}}=-3$ days and after $T_{\rm{rest}}=10$ days when showing the results for all types of simulations. This means that for each candidate we use, there are observations pre- and post-peak epoch.~Since we do not know $T_{\rm{rest}}$ beforehand, we relax the epoch cuts to $T_{\rm{rest}}/(1+z_{\rm{max}})$ as in \citet{kessler2010photo_z}, or observations before $T_{\rm{obs}} = -3/2.2$ days and after $T_{\rm{obs}} = 10/2.2$ since our $z_{\rm{max}} = 1.2$, allowing for more range in $T_{\rm{obs}}$. For the same reason, we use peak MJDs from the `Fmax-clump' method to impose epoch cuts.~Second, we require that there are at least 3 bands with detections of S/N $>$ 8 for realistic simulations.~The number of candidates that remain after selection cuts for the astro-$z$ only results are shown in Table \ref{tab:No_cands_by_selection}.~After imposing the epoch selection cut, we are left with 13,827 or 69.1\% of the SNe Ia, while also imposing the S/N cut leaves us with 8,588 SNe Ia out of the simulated 20,000, or about 42.9\%. 

Lastly, for SN photo-$z$'s and its combinations with the other two redshift estimates where the light-curve fitting method \verb|LCFIT+z| is used, we take the fit probability (FITPROB), calculated by integrating the tail of the $\chi^2$ distribution from the $\chi^2$ of the light-curve fit and the number of degrees of freedom for each of the candidates (\textit{p}-value), and require that $\mathrm{FITPROB} \ge 0.01$.~The number of candidates removed by this selection cut is less than 10\% for all combinations as shown in Table \ref{tab:No_cands_with_SN_photo-z}, but noticeably improves the performance metrics. 

\begin{table}[]
    \centering
    \begin{tabular}{lc}
    \toprule
            Type &  Number of Candidates \\
    \midrule
No selection cut &   20,000 \\
       Epoch cut &   13,827 \\
 Epoch + S/N cut (default) &    8,588 \\
 Epoch + S/N cut + AM cut &    7,528 \\
    \bottomrule
    \end{tabular}    
    \caption{Number of candidates that pass the given selection cut.~Out of the 20,000 simulated candidates, the ones that pass the Epoch cut are used to show the results for Perfect and $\sigma_{\mathrm{syst}}$ only simulations, while the default for realistic simulations is Epoch + S/N cut. The AM cut additionally requires that at least one observation for a candidate has an $\rm{AM} > 1.4$.}
    \label{tab:No_cands_by_selection}
\end{table}

\section{Analysis Methodology} \label{sec:methods}

To measure the astro-$z$ for a given SN, we minimize the $\chi^2$ between the model and \verb|SNANA| simulated DCR shifts. As introduced in Section \ref{sec:data_sims}, we use the extended SALT3 model to calculate the model DCR altitude shifts (MODEL) on a grid of (FILTER, $z$, EPOCH, AM, $x_1$, $c$) with $x_1$ and $c$ being the SALT3 light-curve parameters for stretch and color.~The grids for each of the parameters are: 

\begin{equation} \label{eq:param_grid}
\begin{split}
 z : \text{ } & 0.00 \text{ to } 1.20 \text{, increments of } \Delta z = 0.01 \\ 
 \rm{EPOCH}: \text{ } & -18 \text{ to } 50 \text{, increments of 1 day} \\
\text{AM} : \text{ } & 1.00 \text{ to } 3.00 \text{, increments of 0.01}\\ 
x_1 : \text{ } & -3.0 \text{ to } 2.0 \text{, increments of 0.5} \\ 
c : \text{ } & -0.30 \text{ to } 0.50 \text{, increments of 0.05} 
\end{split}
\end{equation}

\noindent As mentioned earlier in Section \ref{sec:data_sims}, we force the SED to be zero whenever the computed SALT3 fluxes are below zero when calculating the DCR model shifts. These negative fluxes sometimes occur at early phases ($T_{\mathrm{rest}} \le -15$) in the ultraviolet, mostly at extreme color ($|c| \ge 0.2$) and stretch ($|x_1| \ge 2$).~This can sometimes cause an issue, because the DCR shifts are infinite when the denominator of Equation \ref{eq:DCR_alt_shift} is zero.~In such cases, we extrapolate the DCR shifts along the epoch axis using \verb|scipy.interpolate|. 

We measure the DCR shift for simulated data similar to how we would for real data. First, we use the “observed” RA and DEC to compute the altitude of the SN at the telescope site, and subtract the altitude of the ($\sigma_{\rm{tot}}^2$ weighted) \textit{Y}-band average RA and DEC for each candidate. This subtraction is needed because we cannot determine the exact coordinates of the K5V reference star as all the astronomical objects in a given exposure are shifted by DCR as well.~Hence, we take the \textit{Y}-band average coordinates of the SN to be a proxy for the reference star coordinates, as the DCR shifts are smallest in the \textit{Y}-band.~While Poisson noise randomly scatters the \textit{Y}-band positions, the average \textit{Y}-band coordinates among multiple observations is a reasonable approximation.~As only about 12,700 candidates have \textit{Y}-band observations, we use the \textit{z}-band average coordinates instead when we do not have \textit{Y}-band observations for a given SN candidate.~When we use the \textit{z}-band average coordinates as the K5V reference star coordinates, we do not use the \textit{z}-band DCR shifts when constructing our probability distribution functions since the \textit{z}-band average coordinates are sensitive to the DCR shifts in the \textit{z}-band.~Further discussion on the choice of reference star is given in Appendix \ref{Appendix:Detailed_Methodology}.

The $\chi^2$ between the model and the observed altitude shifts are defined as: 
\begin{equation}
    \chi^2 = \sum_i \frac{(\Delta_{\rm{OBS},i} - \Delta_{\rm{MODEL},i})^2}{\sigma_{i, \rm{total}}^2}, 
\end{equation} \label{eq:chi2}
\noindent where OBS indicates a real or simulated observation, $i$ denotes each of the observations for a given SN, and $\sigma_{i, \rm{total}}^2 = \sigma_{i, \rm{stat}}^2 + \sigma_{i, \rm{syst}}^2$ with $\sigma_{i, \rm{stat}}$ given in Equation \ref{eq:sigma_stat} and $\sigma_{i, \rm{syst}} = 5$ mas as stated in Section \ref{sec:data_sims}. After the $\chi^2$ is calculated for a given candidate over the 3-dimensional grid of {FILTER, $x_1$, $c$}, we calculate the posterior: 
\begin{equation}
    \mathcal{P}(z) = \int_{-0.3}^{0.5} \int_{-3.0}^{2.0} P(z, x_1, c) P(x_1)P(c) dx_1 dc 
\end{equation} \label{eq:posterior}

\noindent where $P(z, x_1, c) \propto e^{\frac{-\chi^2}{2}}$ with the $\chi^2$ being the sum over the \textit{ugriz} band-$\chi^2$ (\textit{ugri} when \textit{Y}-band observations are absent) and $P(x_1)$ and $P(c)$ are priors for the light-curve stretch and color.~$P(x_1)$ and $P(c)$ are obtained by binning asymmetric Gaussian distributions with the binning shown in Equation \ref{eq:param_grid}.~The value with highest probability, low-sided and high-sided Gaussian widths are taken to be ($\bar{x}_1$, $\sigma_{x_1, -}$, $\sigma_{x_1, +}$) = ($-0.054$, $0.043$, $0.101$) and ($\bar{c}$, $\sigma_{c, -}$, $\sigma_{c, +}$) = ($0.973$, $1.472$, $0.222$) as in \citet{scolnic2016x1_c}, which are also used as the \verb|SNANA| input.~We do not need to normalize $\mathcal{P}(z)$ since we only need to compute the relative probabilities at each redshift. 


\begin{figure*}
    \centering
    \includegraphics[width=0.96\textwidth]{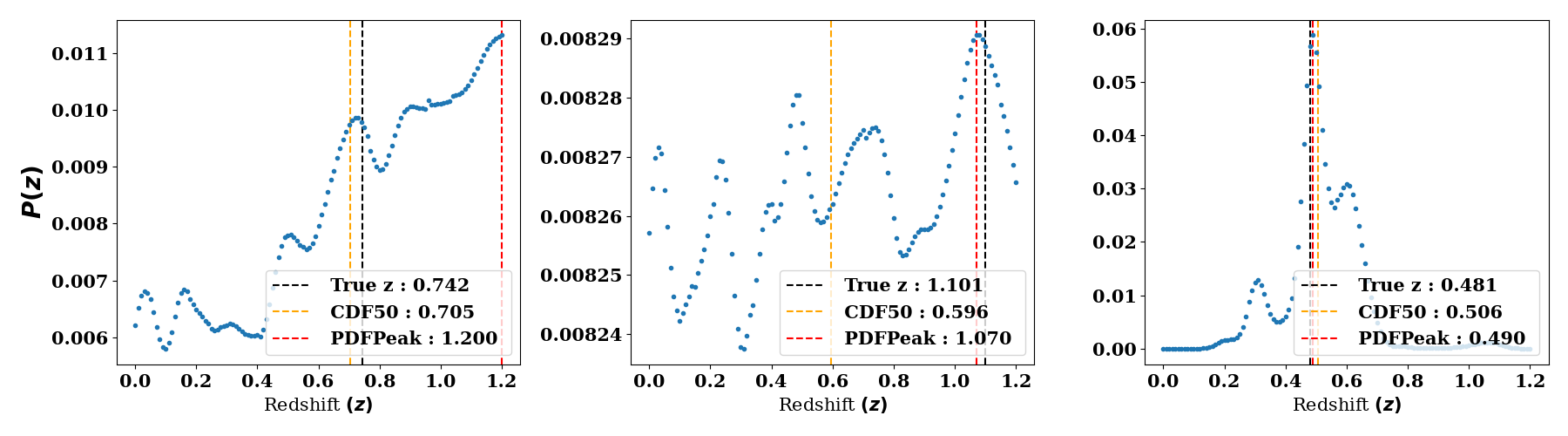}
    \caption{The relative posterior probabilities, $z_{\rm{spec}}$, \CDF{}, and \Peaks{} estimates for three cases highlighting the advantages and disadvantages for the two methods: (Left) \CDF{} gives a better estimate (multimodal with similar peak amplitudes), (Middle) \Peaks{} gives a better estimate (distinct peak but not large in amplitude), and (Right) both \CDF{} and \Peaks{} estimate the correct redshift.}
    \label{fig:posterior_examples}
\end{figure*}

We use $\mathcal{P}(z)$ to estimate the redshift in two different ways:~(1) taking the redshift at the 50th percentile of the cumulative distribution function (hereafter, \CDF{}) and defining the 68\% confidence interval as 16th to 84th percentiles, and (2) taking the redshift at the peak of the probability distribution function (hereafter \Peaks{}, equivalent to the maximum likelihood) with uncertainties defined as the $\pm 34$ percentiles around the peak. 

There are advantages and disadvantages for either method. \CDF{} gives better results when the posterior probabilities are multimodal without a distinct peak and the error bars are always within the range of redshifts we explore.~However, when the $\chi^2$ is unconstraining, the redshift is measured to be near the midpoint of the redshift range, meaning that the results are biased. On the other hand, \Peaks{} can result in catastrophic outliers if the posterior has multimodal peaks or has similar values throughout all redshifts.~Figure \ref{fig:posterior_examples} shows examples for which (i) \CDF{} gives a better redshift estimate (multimodal with similar probabilities), (ii) \Peaks{} gives a better estimate (distinct peak but not large in amplitude), and (iii) both \CDF{} and \Peaks{} are able to determine the redshift very well (well-defined peak).

\begin{table}[]
    \centering
    \begin{tabular}{ll}
    \toprule
            Symbol &  Measurement Method \\
    \midrule
    $z_{\rm{DCR}}$     &     from astrometry using DCR effect\\
    $z_{\rm{Host}}$    &     host photo-$z$ PDF\\
    $z_{\rm{DCR+Host}}$   &  combined PDF from $z_{\rm{DCR}}$ and $z_{\rm{Host}}$    \\
    $z_{\rm{SN}}$     &      from \verb|LCFIT+z|, using flat redshift prior\\
    $z_{\rm{SN+Host}}$   &   from \verb|LCFIT+z|, using $z_{\rm{Host}}$ PDF as prior\\
    $z_{\rm{SN+DCR}}$   &   from \verb|LCFIT+z|, using $z_{\rm{DCR}}$ PDF as prior\\
    $z_{\rm{SN+DCR+Host}}$ & from \verb|LCFIT+z|, using $z_{\rm{DCR+Host}}$ PDF as prior\\
    \bottomrule
    \end{tabular}    
    \caption{How we denote our redshift estimates using various methods and their combinations.}
    \label{tab:measured-z-labels}
\end{table}

In addition to our astro-$z$ results, 
we combine our astrometric redshifts with host-galaxy photometric redshifts (Host photo-$z$'s) as well as SN Ia photometric redshifts (SN photo-$z$'s) in Section \ref{sec:results}.~Henceforth, we denote the redshift measurements using the three methods as well as their combinations as shown in Table \ref{tab:measured-z-labels}.~$z_{\rm{Host}}$ are  
derived from ELAsTiCC host photometry by using the \verb|pzflow| package\footnote{\url{https://github.com/jfcrenshaw/pzflow}} \citep{john_franklin_crenshaw_2021_4679913} as a generative model, trained on the CosmoDC2 simulations \citep{2019ApJS..245...26K},
and are stored as quantiles corresponding to integrated (CDF) probabilities of [$0\%$, $10\%$, ... , $100\%$]. The $z_{\rm{SN}}$'s are obtained from the light-curve fitting method \verb|LCFIT+z| \citep{kessler2010photo_z} using the \textit{ugrizY} bands as with our $z_{\rm{DCR}}$ simulations, where the redshift is fit simultaneously along with the light-curve parameters, i.e., time of maximum brightness $t_0$, color $c$, stretch $x_1$, and flux normalization $x_0$.

When combining $z_{\rm{DCR}}$ with $z_{\rm{Host}}$, we use the \verb|qp| package\footnote{\url{https://github.com/LSSTDESC/qp}} \citep{malz2018_approx_photo-z} to reconstruct PDFs from the $z_{\rm{Host}}$ quantiles before multiplying the $z_{\rm{DCR}}$ and $z_{\rm{Host}}$ PDFs together.~To combine either $z_{\rm{DCR}}$ or $z_{\rm{Host}}$ (or both) with $z_{\rm{SN}}$, we provide the respective quantiles as priors in \verb|LCFIT+z|. We do not determine $P(z_{\rm{DCR}})$, $P(z_{\rm{Host}})$, and $P(z_{\rm{SN}})$ independently and combine the three PDFs for a joint PDF, because the SALT3 parameters such as $x_0$, $x_1$, and $c$ and their uncertainties would be incorrectly associated with $P(z_{\rm{SN}})$ only. Thus the correct method is to combine $P(z_{\rm{Host}})$ and $P(z_{\rm{DCR}})$, and use this combined prior in \verb|LCFIT+z| to not only obtain the combined redshift estimates, but also the SALT3 parameters with correct uncertainties.~When showing combined results, we impose the default selection cut described in Section \ref{sec:data_sims} on the candidates with successful light-curve fits with $z_{\rm{SN}}$ and just the default selection cut when $z_{\rm{SN}}$ are not used.

Sometimes, the $z_{\rm{DCR}}$ and $z_{\rm{Host}}$ estimates cannot be determined (e.g. when the SN candidate has no \textit{z} or \textit{Y} band observations for $z_{\rm{DCR}}$).~In this case, we replace the PDF with a flat prior; if there is no $z_{\rm{DCR}}$ estimate for a candidate, only the $z_{\rm{Host}}$ prior is used and vice versa. 

\begin{figure*}
    \centering
    \includegraphics[width=0.95\textwidth]{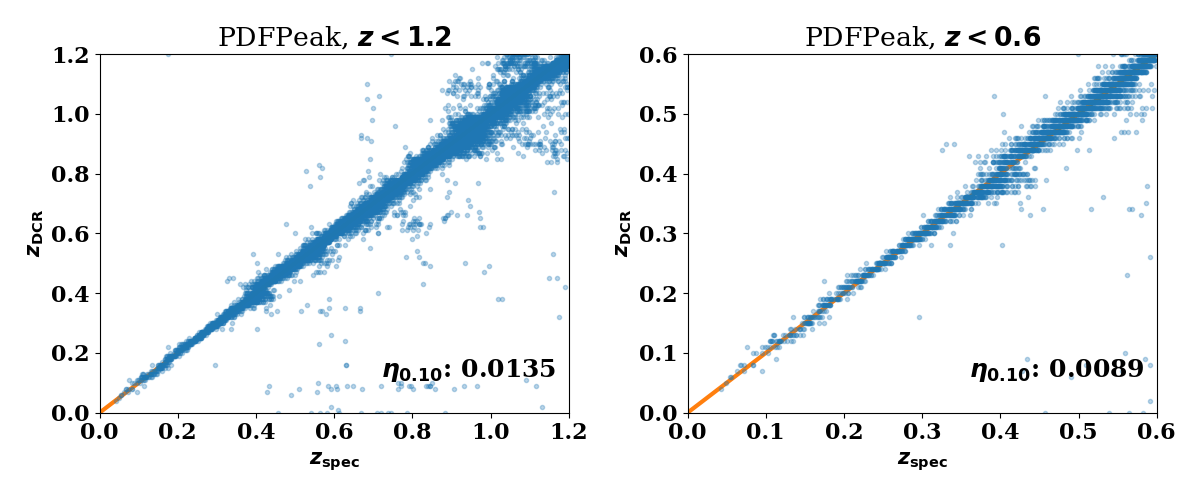}
    \caption{ $z_{\rm{DCR}}$ vs. $z_{\rm{spec}}$ in the Perfect case for \Peaks{}.~As pointed out in Section \ref{sec:methods}, \CDF{} results are equivalent to \Peaks{} results for this case and the $1\sigma$ uncertainties are zero. The right panels are the same as the left, but zoomed to $z<0.6$ and the outlier fraction $\eta_{0.10}$ is defined as the fraction where $|\Delta z| > 0.10$, with $\Delta z \equiv \frac{z_{\rm{estimate}}-z_{\rm{spec}}}{1+z_{\rm{spec}}}$. The imperfect $z_{\rm{DCR}}$ estimates are due to degeneracies in the DCR shifts at several redshifts, which is most apparent at $z_{\rm{spec}} > 0.8$.}
    \label{fig:LSST_z_pred_vs_z_true_ideal}
\end{figure*}

When combining two or more PDFs, sometimes the PDFs have negligible overlap, meaning that it is meaningless to combine them.~Notable cases occur at $z_{\rm true} < 0.4$ where $z_{\rm{DCR}}$ perform better than $z_{\rm{Host}}$; $z_{\rm{DCR+Host}}$ perform worse than $z_{\rm{DCR}}$ by itself when $z_{\rm{DCR}}$ and $z_{\rm{Host}}$ are combined without considering the compatibility between $z_{\rm{DCR}}$ and $z_{\rm{Host}}$ PDFs.~Hence, we impose an additional selection cut for $z_{\rm{DCR+Host}}$ when $z_{\rm{DCR}}$ and $z_{\rm{Host}}$ PDFs are incompatible.~We assess the compatibility between the two PDFs using the Overlapping Index (OVL) between two PDFs, which is defined in the literature such as \citet{inman1989overlapping}, but originally postulated in a related form in \citet{pearson1894}: 
\begin{equation}
    \mathrm{OVL} (p_0, p_1) = \int \min(p_0(x), p_1(x)) dx
\end{equation}

\noindent where $p_0(x)$ and $p_1(x)$ are PDFs. Based on the performance metrics (discussed in the beginning of Section \ref{sec:results}) for $z_{\rm{DCR+Host}}$ compared to $z_{\rm{DCR}}$ and $z_{\rm{Host}}$ by themselves, we remove candidates for $z_{\rm{DCR+Host}}$ when $\mathrm{OVL} < 0.0344$, or when two normalized Gaussian distributions with the same standard deviations ($\sigma$) have means that are $3\sqrt{2} \sigma$ apart from each other. This choice eliminates only about 2.3\% of the candidates that pass the Default selection cut, mostly at the lowest redshifts where incorrect $z_{\rm{Host}}$ PDFs can degrade $z_{\rm{DCR+Host}}$ considerably, but significantly improves performance metrics at those redshifts.~We note that removing candidates using an OVL cut may not be the most optimal way to treat the combination of incompatible PDFs, but is sufficient for this analysis.~While a small OVL can result from distributions with the same mean but very different widths, in our analysis, the use of OVL is (1) limited to a very small number of the candidates and (2) applied when combining with $z_{\mathrm{DCR}}$ and $z_\mathrm{Host}$, neither of which have very narrow PDFs.

The \verb|LCFIT+z| implementation in SNANA produces a $z_{\mathrm{SN}}$ mean and uncertainty, not a PDF, and therefore an OVL value cannot be computed using $z_{\mathrm{SN}}$.~The $\mathrm{FITPROB} \ge 0.01$ cut mentioned in Section \ref{sec:data_sims} rejects most of the incompatible PDFs and we believe that our treatment is sufficient for this initial investigation of combining SN Ia astrometric redshifts with photometric redshifts.~In Appendix \ref{Appendix:Overlap}, we discuss the overlapping index and the treatment of $z_{\rm{SN}}$ PDFs in more detail.

Additionally, for $z_{\rm{SN+DCR+Host}}$, we use either $z_{\rm{DCR}}$ alone or $z_{\rm{Host}}$ alone as priors when $z_{\rm{DCR}}$ and $z_{\rm{Host}}$ are incompatible with each other, depending on whether the overlapping index between $z_{\rm{SN}}$ and $z_{\rm{DCR}}$ or $z_{\rm{Host}}$ is higher.~This choice further improves the performance metrics at low-$z$, as with $z_{\rm{DCR+Host}}$ discussed earlier. 

\section{Results} \label{sec:results}

In this section, we show the results of our $z_{\rm{DCR}}$ estimates for LSST-like simulations for: (i) Perfect ($\sigma_{\rm{stat}} = 0$ and $\sigma_{\rm{syst}}= 0$) (ii) Realistic (with non-zero $\sigma_{\rm{stat}}$ and $\sigma_{\rm{syst}}$).~In Appendix \ref{sec:sigma_syst only}, we show Systematic Effects only.~For the Realistic case, we also show the results combined with $z_{\rm{Host}}$ as well as with $z_{\rm{SN}}$ obtained from \verb|LCFIT+z|. Along with the comparison between the estimated redshifts and true redshifts, we also show three different metrics depending on the case and the combination of PDFs:~the mean bias (or binned residuals), outlier fraction, and the MAD (Median Absolute Deviation) deviation.

As is customary in photo-$z$ literature \citep{kessler2010photo_z,Pasquet2019_SDSS_photoz,qu2023photo_z_ML}, we define the residuals as: 
\begin{equation}
    \Delta z \equiv \frac{z_{\rm{estimate}}-z_{\rm{spec}}}{1+z_{\rm{spec}}}
\end{equation}\label{eq:residual}

\noindent where $z_{\rm{estimate}}$ is the estimated redshift and the bias is $\langle \Delta z \rangle$.~$z_{\rm{spec}}$ refers to the spectroscopic redshift of the SN Ia or its host-galaxy for an analysis using real observations, but is taken to be the true redshift from the simulations in this work.~The uncertainty on $\langle \Delta z \rangle$ is defined as $\frac{\mathrm{RMS}(\Delta z - \langle \Delta z \rangle)}{\sqrt{N_{\mathrm{SN}}}}$ where RMS is the root-mean-square and $N_{\mathrm{SN}}$ is the number of SN events in the sample or redshift bin.~The outlier fraction $\eta_{x}$ is defined as the fraction of candidates where $|\Delta z| > x$:  $x = 0.10$ is our default. Lastly, the precision metric MAD deviation is defined as  $\sigma_{\rm{MAD}} = 1.4826 \times \rm{Median}|\Delta z - \rm{Median}(\Delta z)|$.

\subsection{Perfect Simulations} \label{sec:Perfect}

\begin{figure*}
    \centering
    \includegraphics[width=0.95\textwidth]{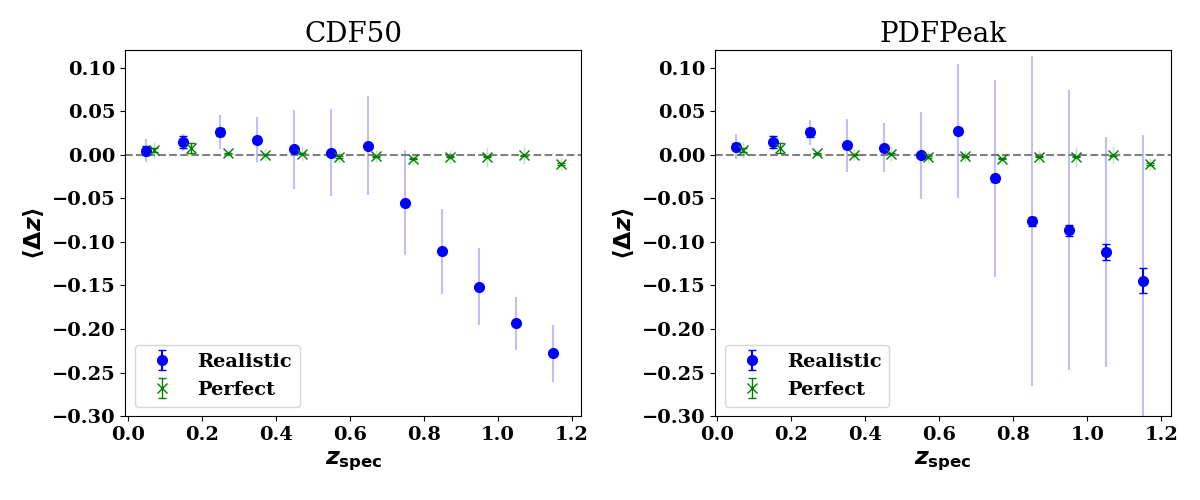}
    \caption{Binned residuals (bias), error on the mean (bold error bars) and the size of $\sigma_{\mathrm{MAD}}$ (light error bars) as a function of $z_{\rm{spec}}$ for Perfect and Realistic simulations for \CDF{} (left) and \Peaks{} (right). \CDF{} and \Peaks{} values are equivalent for Perfect simulations. For clarity, Realistic points (blue) are plotted at the midpoints of each redshift bin ($0.05, 0.15, 0.25, ...$), while Perfect are displaced 0.02 to the right.~The bias for the Realistic case noticeably deviates away from zero starting at $z_{\rm{spec}} = 0.7$, especially for \CDF{}, where $z_{\rm{DCR}} \approx 0.6$ for the unconstrained candidates. \Peaks{}, on the other hand, displays significantly larger $\sigma_{\rm{MAD}}$ at higher $z_{\rm{spec}}$. This is due to $z_{\rm{DCR}}$ being centered at $0.6$ with little spread for \CDF{}, meaning that the redshift estimates are precise, but not accurate.}
    \label{fig:residuals_by_sim}
\end{figure*}

\begin{figure*}
    \centering
    \vspace{-20pt}
    \includegraphics[width=0.95\textwidth]{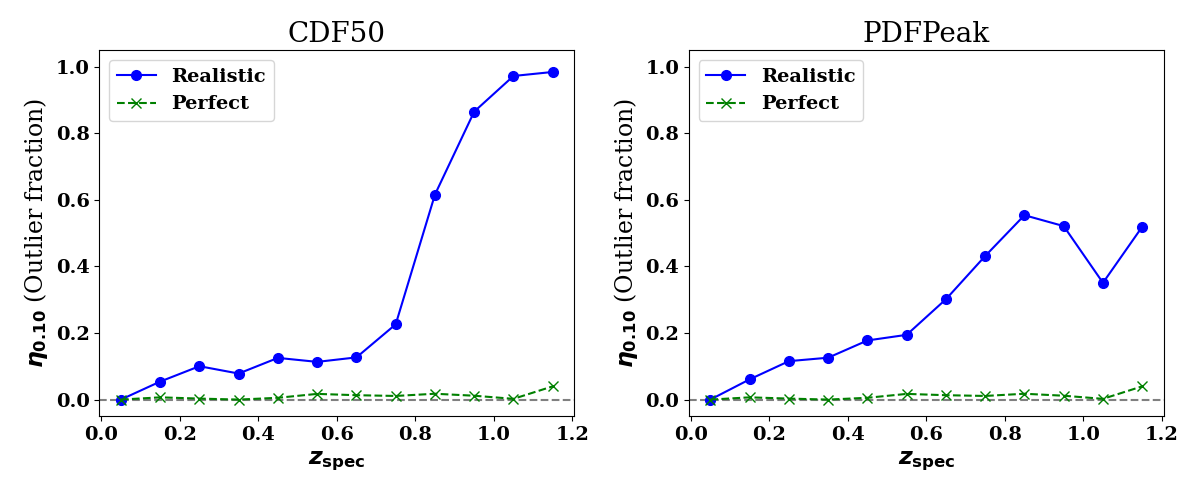}
    \caption{Binned outlier fractions as a function of $z_{\rm{spec}}$ for Perfect and Realistic simulations for \CDF{} (left) and \Peaks{} (right). The Perfect simulations show $ < 5\%$ outlier fractions at all redshifts. For the Realistic simulations, the outlier fractions remain below 40\% for both \CDF{} and \Peaks{} until $z_{\rm{spec}} = 0.7$. At $z_{\rm{spec}} \ge 0.8$, \CDF{} estimates are mostly incorrect as the PDFs are largely flat and unconstrained. Note that the outlier fractions for \Peaks{} decreasing at the highest redshifts for the Realistic case is due to the possible estimated redshift range being limited from 0.0 to 1.2.}
    \label{fig:outliers_rate_by_sim}
\end{figure*}

\begin{table*}[]
    \centering
    \vspace{-10pt}
\begin{tabular}{ccccccc}
\toprule
     Type &            Bias (CDF50) &          Bias (PDFPeak) & $\eta_{0.10}$ (CDF50) & $\eta_{0.10}$ (PDFPeak) & $\sigma_{\rm{MAD}}$ (CDF50) & $\sigma_{\rm{MAD}}$ (PDFPeak) \\
\midrule
  Perfect & -0.003$\pm$0.000 & -0.003$\pm$0.000 &               0.014 &                 0.014 &                     0.008 &                       0.008 \\
Realistic & -0.048$\pm$0.001 & -0.024$\pm$0.002 &               0.333 &                 0.335 &                     0.086 &                       0.077 \\
\bottomrule
\end{tabular}
    \caption{Bias and error on the mean, $\eta_{0.10}$, and $\sigma_{\rm{MAD}}$ for Perfect and Realistic simulations. We show in Section \ref{sec:sigma_syst only} that most of the degradation shown here for the Realistic case is caused by statistical uncertainties.}
    \label{tab:statistics_by_sims}
\end{table*}

We first show our results in Figure \ref{fig:LSST_z_pred_vs_z_true_ideal} for the Perfect case without any statistical or systematic uncertainties, assuming that our peak MJD measurements have no error, and that we know both $x_1$ and $c$. This test validates the analysis and demonstrates the ideal performance of $z_{\rm{DCR}}$ for the LSST DDF SNe Ia.~Only the epoch selection cut is applied, since we are essentially assuming infinite S/N. As highlighted in Section \ref{sec:methods}, we set $\sigma_{\mathrm{syst}} = 1$ mas in the $\chi^2$ denominator to avoid infinite $\chi^2$ (Equation \ref{eq:chi2}), while $\sigma_{\mathrm{syst}} = 0$ in the simulation.~The results we show here are equivalent for \Peaks{} and \CDF{}.

In this perfect case, $z_{\rm{DCR}}$ give very precise estimates of the SN Ia redshifts for all redshifts, but especially for $z < 0.6$ where $z_{\rm{DCR}}$ almost always coincides with $z_{\rm{spec}}$. When all redshifts are included, the outlier fraction $\eta_{0.10}$ is 1.35\%, $\sigma_{\rm{MAD}}$ is 0.008, and bias is $-0.003 \pm 0.000$ as shown in Table \ref{tab:statistics_by_sims}.

\begin{figure*}
    \vspace{-30pt}
    \centering
    \includegraphics[width=0.96\textwidth]{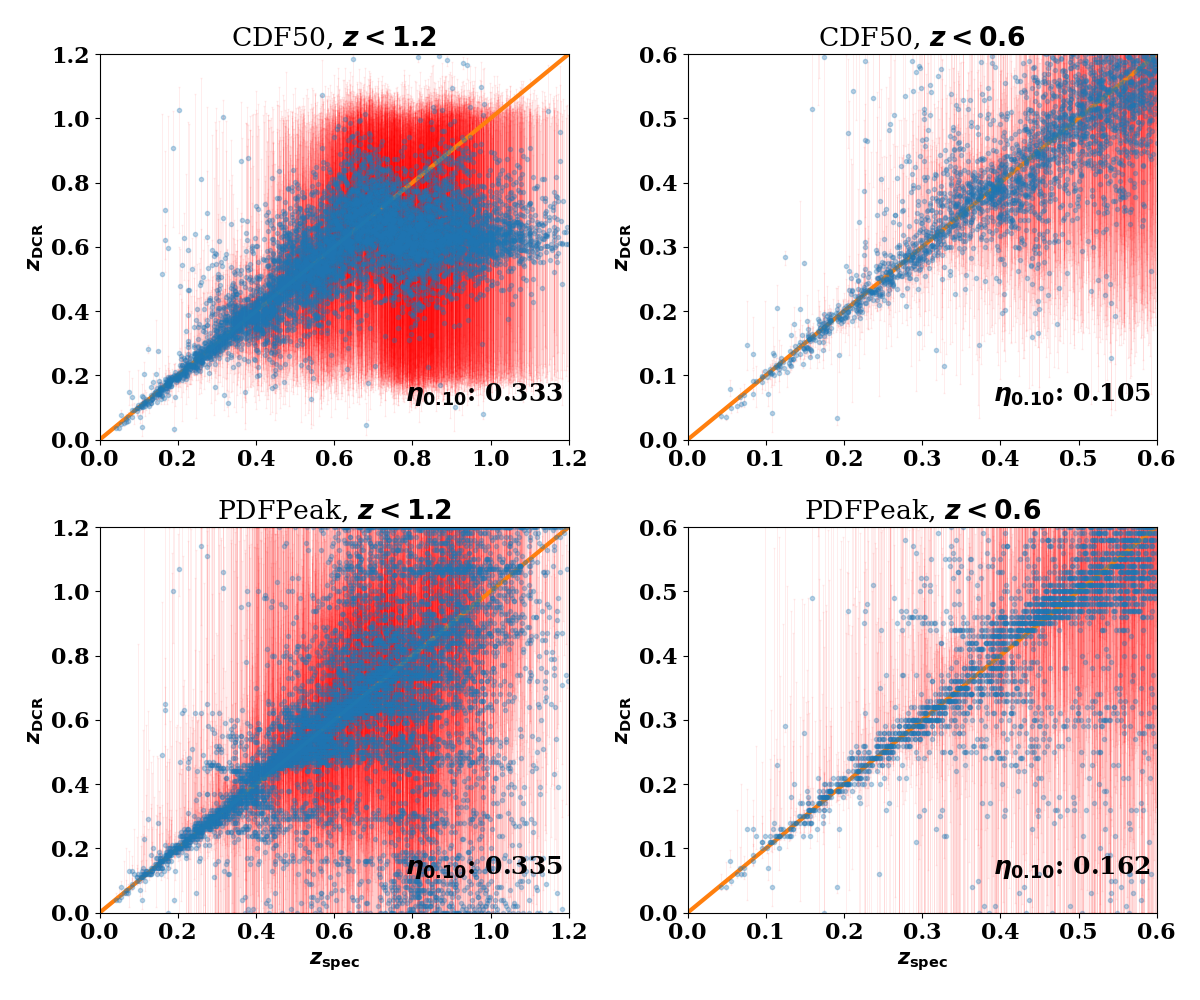}
    \caption{Similar to Figure \ref{fig:LSST_z_pred_vs_z_true_ideal}, but for the Realistic case; $z_{\rm{DCR}}$ vs. $z_{\rm{spec}}$ and the $1\sigma$ error bars for the Realistic case for \CDF{} (top panels) and \Peaks{} (bottom panels). We see noticeable degradation in the $z_{\rm{DCR}}$ estimates, especially at $z_{\rm{spec}} > 0.8$. The $z_{\rm{spec}} < 0.6$ panels still show good agreement.}
    \label{fig:realistic_z_pred_vs_z_true}
\end{figure*}

In Figure \ref{fig:residuals_by_sim} and Figure \ref{fig:outliers_rate_by_sim}, the green X symbols show the binned bias, $\sigma_{\mathrm{MAD}}$, and the outlier fractions versus $z_{\rm{spec}}$. Again, the values shown for \CDF{} and \Peaks{} are equivalent. For the Perfect case, the bias is below 0.01 and $\eta_{0.10}$ is also below 5\% in all redshift bins up to $z_{\rm{spec}} = 1.2$. 

Although the results shown here are for the Perfect case, there are some degeneracies and catastrophic outliers in Figure \ref{fig:LSST_z_pred_vs_z_true_ideal} (clumps or streaks) especially at the high redshifts.~These artifacts are caused by the DCR shifts being degenerate at multiple redshifts at a given AM and observed epoch; fewer observations result in a higher probability that these degeneracies remain.~Additionally, at higher redshifts, we typically do not have detections in the \textit{ug} bands due to low S/N (we only use detections where the coadded $\mathrm{S/N} > 3$ as mentioned in Section \ref{sec:data_sims}) where the DCR shifts are largest and therefore most informative.

\begin{figure*}
    \centering
    \vspace{-15pt}
    \includegraphics[width=0.96\textwidth]{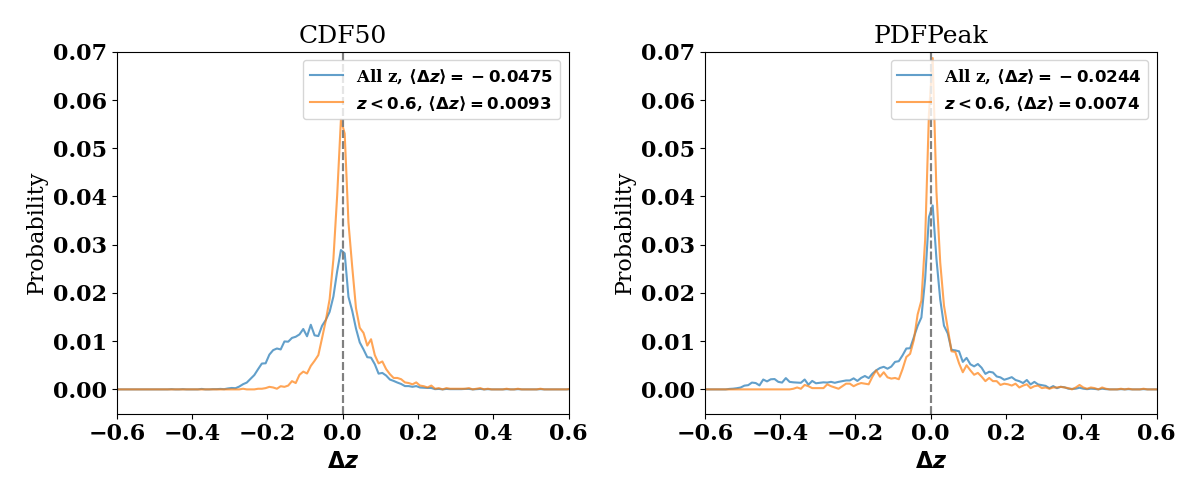}
    \caption{Histograms of the residuals $\Delta z$ for the Realistic case for \CDF{} (left) and \Peaks{} (right). While the overall bias is significantly larger than the Perfect case, we see little bias at $z_{\rm{spec}} < 0.6$.}
    \label{fig:realistic_histograms}
\end{figure*}


\subsection{Realistic Simulations}\label{sec:realistic}

In Figures \ref{fig:realistic_z_pred_vs_z_true} and \ref{fig:realistic_histograms}, we present our results for the realistic case with both $\sigma_{\rm{stat}}$, $\sigma_{\rm{syst}}$, and realistic (mostly light-curve fit) peak MJD measurements.~Additionally, we marginalize over $x_1$ and $c$.~After imposing the S/N selection cut on top of the epoch selection cut, we are left with about 42.9\% of the simulated candidates (see Section \ref{sec:data_sims}).

Here, we find that the $z_{\rm{DCR}}$ estimates have degraded compared to the Perfect case, especially at high redshifts ($z_{\rm{spec}} > 0.7$).~\CDF{} estimates very few SNe to be above $z = 0.8$ because the PDFs are nearly flat and the 50th percentile is clustered near the middle of the redshift range as shown in Figure \ref{fig:realistic_z_pred_vs_z_true} and reflected in the histograms of the residuals (Figure \ref{fig:realistic_histograms}).~\Peaks{} and \CDF{} show similar performance when all redshifts are included in terms of $\eta_{0.10}$ (33.3\% and 33.5\% respectively), while the bias is lower for \Peaks{} ($-0.024 \pm 0.002$ compared to $-0.048 \pm 0.001$ for \CDF{}).
This suggests that even at $z_{\rm{spec}} > 0.8$, the high redshift PDF peaks still provide some information, although by a modest amount.


In the $z_{\rm{spec}} < 0.6$ panels of Figure \ref{fig:realistic_z_pred_vs_z_true}, the accuracy is still relatively high compared to $z_{\rm{spec}} > 0.6$, but \CDF{} shows better performance. This is because \CDF{} estimates the redshift to be $z \approx 0.6$ when the PDF is unconstraining, reducing catastrophic outliers when $z_{\rm{spec}} < 0.6$. In Figure \ref{fig:residuals_by_sim} and Figure \ref{fig:outliers_rate_by_sim}, the biases are close to zero at $z_{\rm{spec}} < 0.7$ and the outlier fractions are below 20\% at $z_{\rm{spec}} < 0.6$ for both \CDF{} and \Peaks{}. The bias however, deviates from zero at $z_{\rm{spec}} > 0.7$ for both \CDF{} and \Peaks{}, but to a greater extent for \CDF{} since most of the estimated redshifts are around $0.6$.~This is also the reason why the $\sigma_{\rm{MAD}}$ error bars are much larger for \Peaks{} compared to \CDF{}; the \CDF{} estimates are precise, but not accurate at high-$z$. The diagonal streak (Figure \ref{fig:realistic_z_pred_vs_z_true}) spanning $(z_{\rm{spec}}, z_{\rm{estimate}})$ = $(0.3, 0.5)$ to $(0.5, 0.3)$ could be due to the \textit{g}-band DCR shifts typically being close to zero throughout all epochs around these redshifts, while there likely are not many \textit{u}-band observations, increasing room for degeneracies. In Appendix \ref{sec:realistic_cuts}, we further discuss the sensitivity of our results on various selection cuts. 

\begin{figure*}
    \centering
    \includegraphics[width=0.96\textwidth]{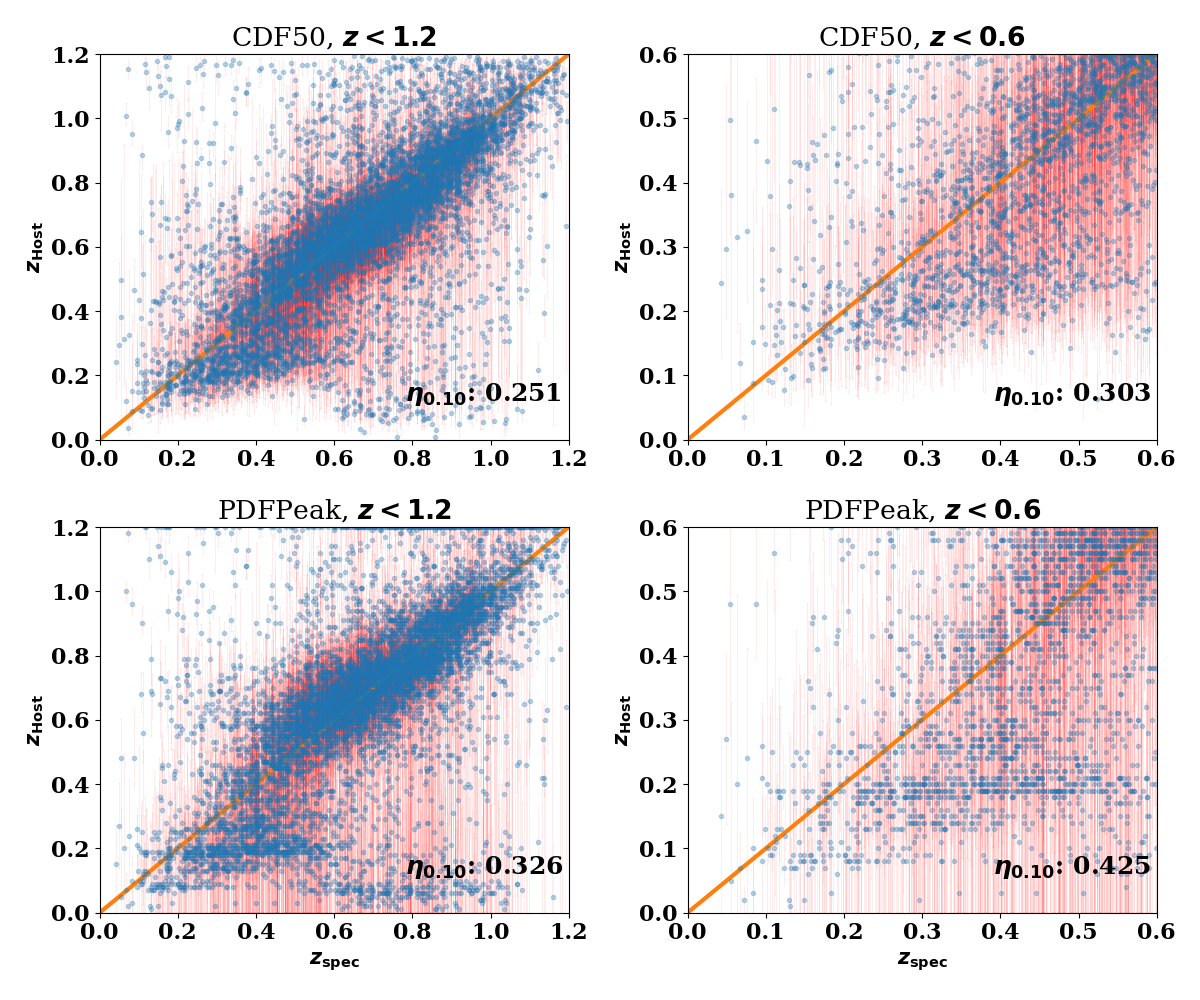}
    \caption{$z_{\rm{Host}}$ vs. $z_{\rm{spec}}$ and the $1\sigma$ error bars for \CDF{} (top panels) and \Peaks{} (bottom panels).~For around 5\% of the SNe that passes the Default selection cut, there are no $z_{\rm{Host}}$ estimates, so we assume flat PDFs as discussed in Section \ref{sec:methods}. While we do not show these candidates as data points and error bars in this figure for clarity, they are always included for the performance metrics.~We see that the $z_{\rm{Host}}$ is more accurate at higher redshifts contrary to $z_{\rm{DCR}}$.~We also note that \CDF{} performs better here, although this could be the result of the PDFs being constructed from quantiles with 11 bins.}
    \label{fig:host_photo-z_only}
\end{figure*}

\begin{table*}[]
    \centering
\begin{tabular}{ccccccc}
\toprule
               Type &            Bias (CDF50) &          Bias (PDFPeak) & $\eta_{0.10}$ (CDF50) & $\eta_{0.10}$ (PDFPeak) & $\sigma_{\rm{MAD}}$ (CDF50) & $\sigma_{\rm{MAD}}$ (PDFPeak) \\
\midrule
     $z_{\rm{DCR}}$ & -0.048$\pm$0.001 & -0.024$\pm$0.002 &               0.333 &                 0.335 &                     0.086 &                       0.077 \\
    $z_{\rm{Host}}$ & -0.010$\pm$0.001 & -0.032$\pm$0.002 &               0.251 &                 0.326 &                     0.067 &                       0.085 \\
$z_{\rm{DCR+Host}}$ & -0.016$\pm$0.001 & -0.013$\pm$0.001 &               0.156 &                 0.189 &                     0.047 &                       0.051 \\
\bottomrule
\end{tabular}
    \caption{Bias and error on the mean, $\eta_{0.10}$, and $\sigma_{\rm{MAD}}$ for $z_{\rm{DCR}}$ only, $z_{\rm{Host}}$ only, and $z_{\rm{DCR+Host}}$. Combining the two reduces the bias (\Peaks{}) and $\sigma_{\rm{MAD}}$, while the outlier fractions are reduced by at least 10\%.}
    \label{tab:statistics_with_host_z}
\end{table*}

\begin{figure*}
    \vspace{-43pt}
    \centering
    \includegraphics[width=0.95\textwidth]{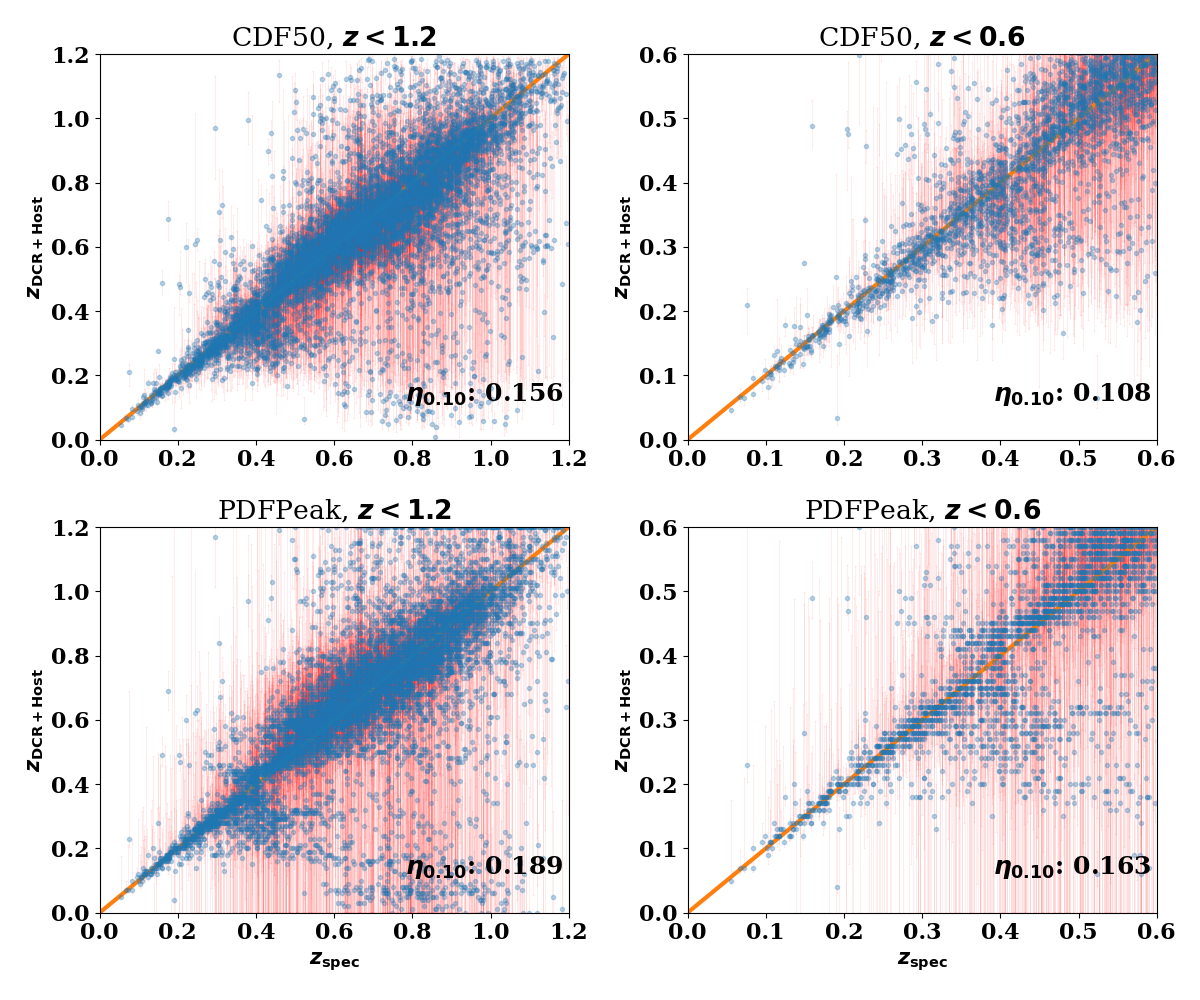}
    \caption{$z_{\rm{DCR+Host}}$ vs. $z_{\rm{spec}}$ and the $1\sigma$ error bars for \CDF{} (top panels) and \Peaks{} (bottom panels). We see noticeable improvement from Figure \ref{fig:host_photo-z_only}, especially at $z_{\rm{spec}} < 0.6$ where the $z_{\rm{Host}}$ are not very constraining.}
    \label{fig:host_photo-z+astro-z}
\end{figure*}

\begin{figure*}
    \vspace{-10pt}
    \centering
    \includegraphics[width=0.96\textwidth]{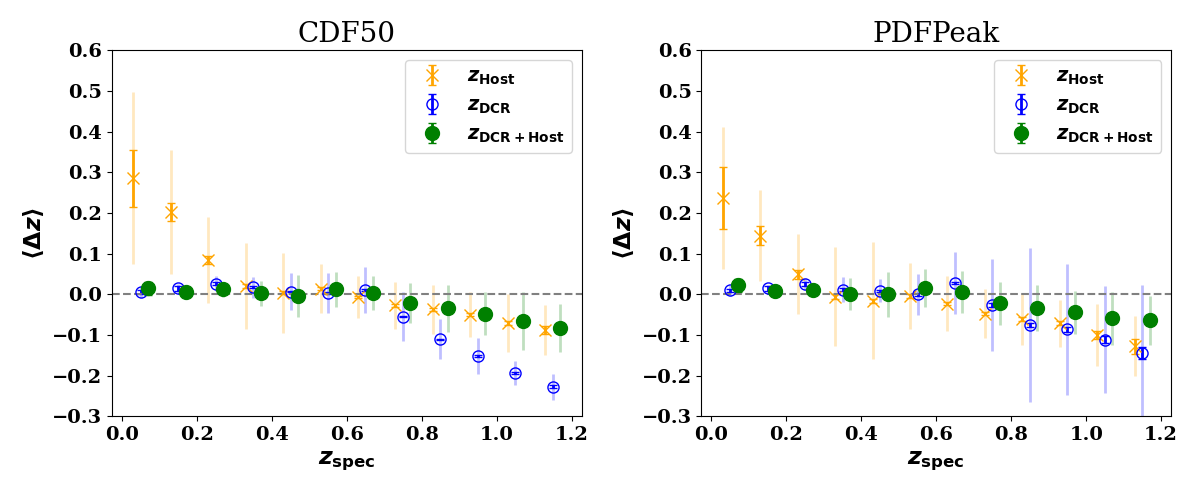}
    \caption{Binned residuals (bias), error on the mean (bold error bars) and the size of $\sigma_{\mathrm{MAD}}$ (light error bars) as a function of $z_{\rm{spec}}$ for $z_{\rm{DCR}}$, $z_{\rm{Host}}$, and $z_{\rm{DCR+Host}}$ for \CDF{} (left) and \Peaks{} (right). Again, $z_{\rm{DCR}}$ is placed at the midpoints of each redshift bin, while $z_{\rm{Host}}$ and $z_{\rm{DCR+Host}}$ are displaced to the left and right respectively. Contrary to $z_{\rm{DCR}}$, $z_{\rm{Host}}$ display more bias and larger error bars at lower redshifts, but combining with $z_{\rm{DCR}}$ alleviates this.}
    \label{fig:residuals_host_photo-z_combined}
\end{figure*}

\subsubsection{Combining with $z_{\rm{Host}}$}\label{sec:astro-z+host_photo-z}

\begin{figure*}
    \centering
    \includegraphics[width=0.96\textwidth]{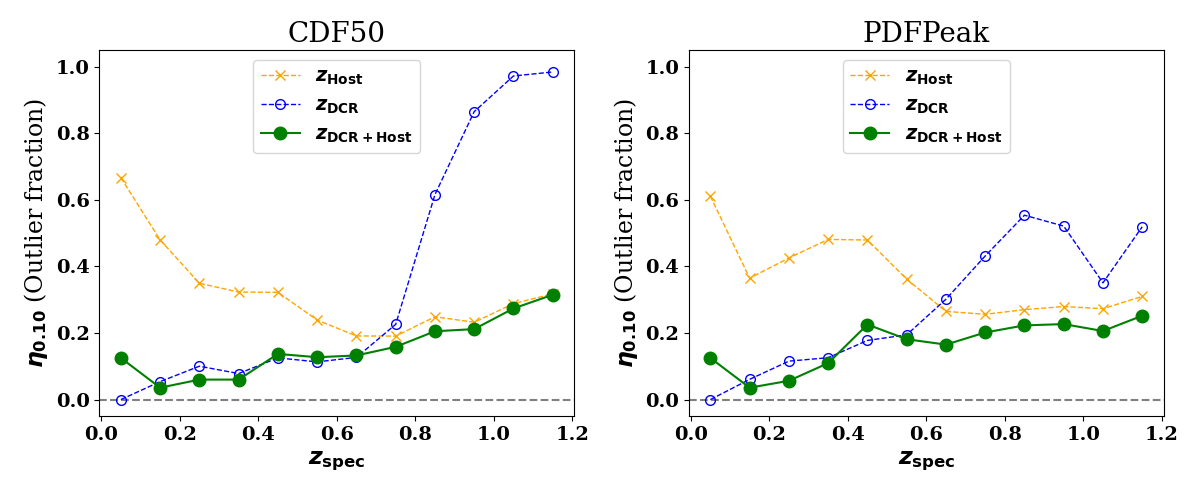}
    \caption{Binned outlier fractions as a function of $z_{\rm{spec}}$ for $z_{\rm{DCR}}$, $z_{\rm{Host}}$, and $z_{\rm{DCR+Host}}$  for \CDF{} (left) and \Peaks{} (right). Combining with $z_{\rm{DCR}}$ significantly reduces the outlier fractions of $z_{\rm{Host}}$ at $z_{\rm{spec}} < 0.6$ for both \CDF{} and \Peaks{}, while the high-$z$ outlier fractions are not affected by much.}
    \label{fig:outlier_rates_host_photo-z_combined}
\end{figure*}

In this section, we discuss the performance of host-galaxy photometric redshifts ($z_{\rm{Host}}$) combined with astrometric redshifts. Since we estimate $z_{\rm{Host}}$ using PDFs constructed from quantiles, we apply the \CDF{} and \Peaks{} methods the same way as for $z_{\rm{DCR}}$.~First, in Figure \ref{fig:host_photo-z_only}, we show $z_{\rm{Host}}$ as a function of $z_{\rm{spec}}$ for \CDF{} and \Peaks{}, with the same default selection cut imposed for the $z_{\rm{DCR}}$ only simulations. Compared to $z_{\rm{DCR}}$, the $z_{\rm{Host}}$ are much more accurate at higher redshifts, but less accurate at lower redshifts. This is due to the distinct 4000\AA{} break for galaxies being in the \textit{u}-band at low redshifts, where the sensitivity is typically poor. Also, \CDF{} gives more accurate estimates than \Peaks{} with $\eta_{0.10} = 25.1\%$ for \CDF{} and $\eta_{0.10} = 32.6\%$ for \Peaks{}.~As shown in Table \ref{tab:statistics_with_host_z}, \CDF{} performs better than \Peaks{} in terms of bias and $\sigma_{\rm{MAD}}$ as well.

In Figure \ref{fig:host_photo-z+astro-z}, we show the combined results, $z_{\rm{DCR+Host}}$ vs.~$z_{\rm{spec}}$.~The estimated redshifts more tightly follow the true redshifts with $\eta_{0.10}$ being 15.6\% for \CDF{} and 18.9\% for \Peaks{}, meaning that the outlier fractions are reduced by at least 10\% for both \CDF{} and \Peaks{} compared to $z_{\rm{Host}}$.~The improvement is particularly striking at $z_{\rm{spec}} < 0.6$, where the $z_{\rm{Host}}$ are not very constraining.

In Table \ref{tab:statistics_with_host_z}, we also highlight the bias and $\sigma_{\rm{MAD}}$ for $z_{\rm{DCR}}$, $z_{\rm{Host}}$, and $z_{\rm{DCR+Host}}$. We find that combining the two removes most of the bias for \Peaks{} and $\sigma_{\rm{MAD}}$ is considerably reduced for both \CDF{} and \Peaks{}. 

Figures \ref{fig:residuals_host_photo-z_combined}-\ref{fig:outlier_rates_host_photo-z_combined} show the mean binned residuals vs $z_{\rm{spec}}$. We see large bias for $z_{\rm{Host}}$ at low redshifts for both \CDF{} and \Peaks{}, little bias at the mid redshifts, and some bias at the higher redshifts. Combining with $z_{\rm{DCR}}$ reduces the bias, especially at the low redshifts. We note a similar trend for the outlier fractions as well; the $z_{\rm{Host}}$ outlier fractions are reduced significantly when combined with $z_{\rm{DCR}}$ at low redshifts. 

\subsubsection{Combining with $z_{\rm{SN}}$}\label{sec:zSN+DCR+Host}

\begin{table*}
    \centering
    \begin{tabular}{cccc}
    \toprule
     & \multicolumn{3}{c}{Number of LCFIT+z events with:} \\
                Type &       No cuts & Default cuts & Default + FITPROB cuts \\
    \midrule
          $z_{\rm{SN}}$ &  12,537 & 7,206 & 6,577 \\
    $z_{\rm{SN+Host}}$ &  12,397 & 7,134 & 6,585
 \\
$z_{\rm{SN+DCR}}$ &  12,230 & 7,596 & 7,224
 \\
        $z_{\rm{SN+DCR+Host}}$ &  12,845 & 7,534 & 7,227
 \\
    \bottomrule
    \end{tabular}
    \caption{Number of candidates with converged LCFIT+z light-curves (No cuts), with the default selection cut imposed on the LCFIT+z converged candidates (Default cuts), and with both the default selection cut and FITPROB $\ge 0.01$ cut imposed (Default + FITPROB cuts). Combining $z_{\rm{DCR}}$ recovers about 600 more candidates (9\%) where LCFIT+z for photo-$z$'s only do not converge.}
    \label{tab:No_cands_with_SN_photo-z}
\end{table*}

In this section, we show the redshift estimates for supernovae photometric redshifts ($z_{\rm{SN}}$) using \verb|LCFIT+z| as well as $z_{\rm{SN}}$ combined with $z_{\rm{Host}}$ and $z_{\rm{DCR}}$. Assessing the improvement resulting from combining the $z_{\rm{DCR}}$ with  $z_{\rm{SN+Host}}$ is crucial to understanding how $z_{\rm{DCR}}$ can be used for upcoming surveys such as Rubin where the number of supernovae detected will be far too large for spectroscopic redshift measurements. 

\begin{figure*}
    \centering
    \includegraphics[width=0.91
\textwidth]{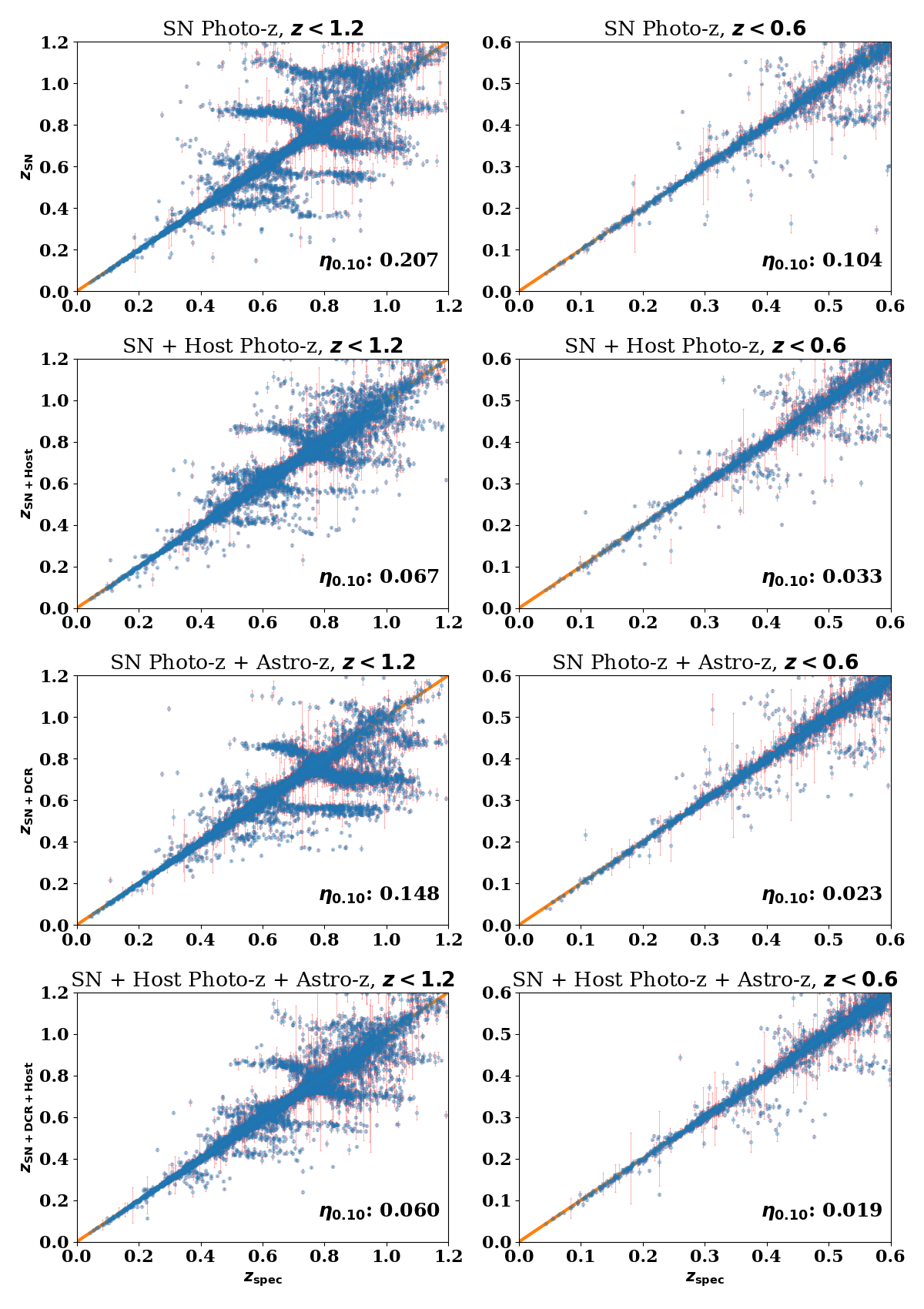}
    \caption{$z_{\rm{SN}}$ vs. $z_{\rm{spec}}$ (top panels), $z_{\rm{SN+Host}}$ vs. $z_{\rm{spec}}$ (second panels), $z_{\rm{SN+DCR}}$ vs. $z_{\rm{spec}}$ (third panels), and $z_{\rm{SN+DCR+Host}}$ vs. $z_{\rm{spec}}$ (bottom panels). Combining $z_{\rm{SN}}$ with either $z_{\rm{Host}}$ or $z_{\rm{DCR}}$ significantly improves the estimates, more so at $z_{\rm{spec}} > 0.6$ for $z_{\rm{Host}}$ and $z_{\rm{spec}} < 0.6$ for $z_{\rm{DCR}}$. Combining all three shows much better overall (throughout all redshifts) accuracy than any two combinations.}
    \label{fig:SN_photo-z_combined}
\end{figure*}

We show in Table \ref{tab:No_cands_with_SN_photo-z}, the number of candidates remaining from the original 20,000 after requiring \verb|LCFIT+z| convergence which is about 61 to 64\% of the candidates, and then the default and FITPROB selection cuts, leaving 33\% to 36\% of the candidates. In Figure \ref{fig:SN_photo-z_combined}, we show $z_{\rm{SN}}$ vs. $z_{\rm{spec}}$ (top panels), $z_{\rm{SN+Host}}$ vs. $z_{\rm{spec}}$ (second panels), $z_{\rm{SN+DCR}}$ vs. $z_{\rm{spec}}$ (third panels), and $z_{\rm{SN+DCR+Host}}$ vs. $z_{\rm{spec}}$ (bottom panels) with the \verb|LCFIT+z| + Default selection cut. We would like to emphasize that including $z_{\mathrm{DCR}}$ to the photo-$z$'s results in a 10\% increase in the number of candidates that pass the Default + FITPROB cuts.

As with $z_{\rm{DCR}}$, $z_{\rm{SN}}$ have streaks of degeneracies at several locations and is more accurate at lower redshifts, although the performance is generally better with $\eta_{0.10}$ being 20.7\%for all redshifts and 10.4\% when limited to $z_{\rm{spec}} < 0.6$. Combining with $z_{\rm{Host}}$ significantly improves the estimates at all redshifts, but the improvement is more pronounced at $z_{\rm{spec}} > 0.6$, with $\eta_{0.10}$ reduced by about 14\% for the all-$z$ sample.~This improvement is expected since $z_{\rm{SN}}$ and $z_{\rm{Host}}$ are independent estimates (as are $z_{\rm{DCR}}$ with $z_{\rm{Host}}$ as discussed in Section \ref{sec:astro-z+host_photo-z}). The third panels show that adding $z_{\rm{DCR}}$ to $z_{\rm{SN}}$ considerably improves the estimates with $\eta_{0.10}$ reduced by about 6\% for the all-$z$ sample compared to $z_{\rm{SN}}$ only. Contrary to combining $z_{\rm{Host}}$, the $z_{\mathrm{SN+DCR}}$ improvement is more prominent at lower redshifts with $\eta_{0.10}$ reduced by 8\% when limited to $z_{\rm{spec}} < 0.6$. Lastly, combining $z_{\rm{SN}}$ with $z_{\rm{Host}}$ and $z_{\rm{DCR}}$ shows the full potential of non-spectroscopic redshift estimates, with $\eta_{0.10}$ decreasing to 6.0\% for all redshifts and 1.9\% when limited to $z_{\rm{spec}} < 0.6$. 

Table \ref{tab:statistics_with_SN_photo-z} shows the bias, $\eta_{0.10}$, and $\sigma_{\rm{MAD}}$ for the four combinations of redshift estimation methods involving $z_{\rm{SN}}$.~The measurements involving $z_{\rm{SN}}$ are unbiased, with the exception of $z_{\mathrm{SN+DCR}}$. Combining $z_{\rm{Host}}$ with $z_{\rm{SN}}$ reduces $\sigma_{\rm{MAD}}$ to about two thirds of the $z_{\rm{SN}}$-only value, while combining $z_{\rm{DCR}}$ with $z_{\rm{SN}}$ shows a similar impact.~When all three methods are combined, the bias is nearly zero and $\sigma_{\rm{MAD}}$ is about 0.014.

\begin{table}[h]
    \centering
\begin{tabular}{cccc}
\toprule
                  Type &                    Bias & $\eta_{0.10}$ & $\sigma_{\rm{MAD}}$ \\
\midrule
         $z_{\rm{SN}}$ &  0.007$\pm$0.001 &         0.207 &               0.026 \\
    $z_{\rm{SN+Host}}$ & -0.000$\pm$0.001 &         0.067 &               0.016 \\
     $z_{\rm{SN+DCR}}$ & -0.019$\pm$0.001 &         0.148 &               0.017 \\
$z_{\rm{SN+DCR+Host}}$ & -0.003$\pm$0.001 &         0.060 &               0.014 \\
\bottomrule
\end{tabular}
    \caption{Bias and error on the mean, $\eta_{0.10}$, and $\sigma_{\rm{MAD}}$ for $z_{\rm{SN}}$, $z_{\rm{SN+Host}}$, $z_{\rm{SN+DCR}}$, and $z_{\rm{SN+DCR+Host}}$. Combining all three methods significantly lowers the outlier fraction compared to $z_{\rm{SN}}$ only.}
    \label{tab:statistics_with_SN_photo-z}
\end{table}

There are a few points worth noting. First, while combining $z_{\rm{SN}}$ with $z_{\rm{Host}}$ reduces the outlier fraction by 14\% as opposed to 6\% with the addition of $z_{\rm{DCR}}$, only about 37\% of the true redshifts (for $z_{\mathrm{SN+DCR+Host}}$) are below 0.6, where the $z_{\rm{DCR}}$ are more informative. Because the $z_{\rm{DCR}}$ and $z_{\rm{Host}}$ each perform better at different redshifts, we show in Table \ref{tab:statistics_low_high_z} the bias, $\eta_{0.10}$, and $\sigma_{\rm{MAD}}$ for the different combinations at high-$z$ ($z_{\rm{spec}} > 0.6$) and low-$z$ ($z_{\rm{spec}} < 0.6$) separately.~For $z_{\rm{DCR}}$, $z_{\rm{Host}}$, and $z_{\rm{DCR+Host}}$, we show the \Peaks{} values.~Second, it may be counter-intuitive that combining $z_{\rm{SN}}$ with $z_{\rm{DCR}}$ results in any noticeable improvement, since the information for both redshift estimates is based on
the same SN Ia SED time series. We evaluate the independence of these two redshift estimates by computing the Pearson correlation coefficient between $z_{\rm{SN}}-z_{\rm{spec}}$ and $z_{\rm{DCR}}-z_{\rm{spec}}$, which we find to be 0.121.~Such a low value shows that the two measurements are largely independent.~This is  because $z_{\rm{DCR}}$ relies on the product of the astrometric shifts due to DCR and the SED within a band (Equation~\ref{eq:DCR_alt_shift}) while $z_{\rm{SN}}$ relies on only the SED integrated within a band. It is therefore reasonable that combining these two measurements results in a better redshift constraint.

Because $z_{\rm{DCR}}$ perform better at low-$z$ compared to $z_{\rm{Host}}$ and vice versa, we discuss our results in terms of low-$z$ and high-$z$ separately in Appendix \ref{sec:zSN+DCR+Host_low-z_high-z}.

\begin{figure}
    \centering
    \includegraphics[width=0.48\textwidth]{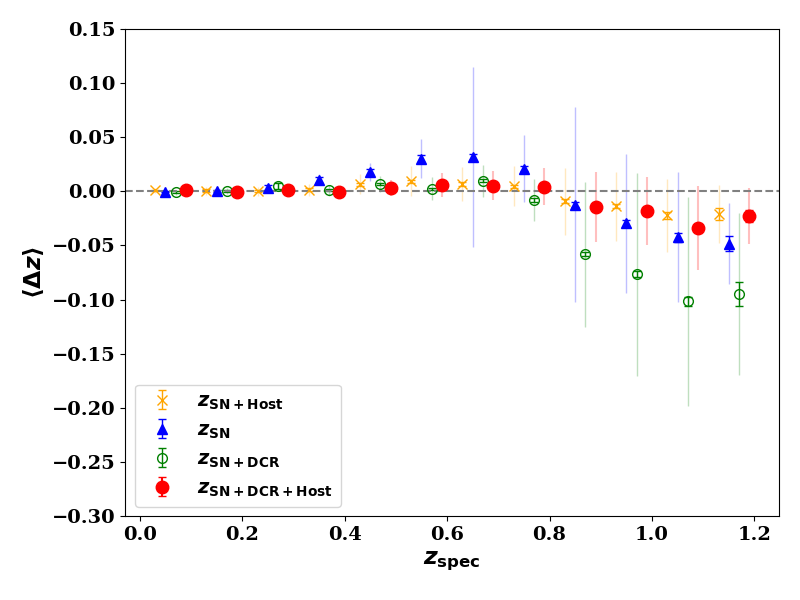}
    \caption{Binned residuals (bias), error on the mean (bold error bars) and the size of $\sigma_{\mathrm{MAD}}$ (light error bars) as a function of $z_{\rm{spec}}$ for $z_{\rm{SN}}$, $z_{\rm{SN+Host}}$ , $z_{\rm{SN+DCR}}$, and $z_{\rm{SN+DCR+Host}}$. Combining all three reduces the bias throughout all redshifts.}
    \label{fig:residuals_with_SN_photo-z}
\end{figure}

\begin{figure}
    \centering
    \includegraphics[width=0.48\textwidth]{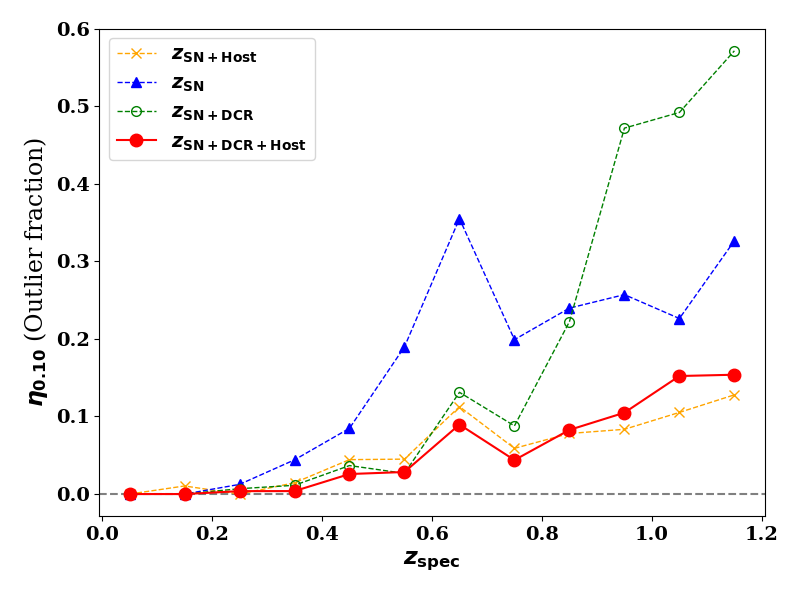}
    \caption{Binned $\eta_{0.10}$ as a function of $z_{\rm{spec}}$ for $z_{\rm{SN}}$, $z_{\rm{SN+Host}}$ , $z_{\rm{SN+DCR}}$, and $z_{\rm{SN+DCR+Host}}$. Combining all three lowers the outlier fractions throughout all redshifts, although $z_{\rm{SN+Host}}$ show slightly lower outlier fractions at $z_{\rm{spec}} > 0.8$.}
    \label{fig:outlier_rates_with_SN_photo-z}
\end{figure}

Lastly, in Figure \ref{fig:residuals_with_SN_photo-z} and Figure \ref{fig:outlier_rates_with_SN_photo-z}, we show the mean binned residuals and its $\sigma_{\rm{MAD}}$ error bars, as well as the outlier fractions by true redshift. Generally speaking, the smallest bias occurs at $z_{\rm{spec}} < 0.2$ for $z_{\rm{SN+DCR}}$, $0.2 <z_{\rm{spec}} < 0.8$ for $z_{\rm{SN+DCR+Host}}$, and $z_{\rm{spec}} > 0.8$ for $z_{\rm{SN+Host}}$. Similarly, the lowest outlier fractions occur at $z_{\rm{spec}} < 0.8$ for $z_{\rm{SN+DCR+Host}}$ and $z_{\rm{spec}} > 0.8$ for $z_{\rm{SN+Host}}$.~For $z_{\rm{SN+DCR+Host}}$, the outlier fractions are lowest at $ 0.0 < z_{\rm{spec}} < 0.2$ (close to 0) and highest at $ 1.1 < z_{\rm{spec}} < 1.2$ (still less than 20\%). Contrary to when $z_{\rm{DCR}}$ are combined with $z_{\rm{Host}}$, combining $z_{\rm{DCR}}$ with $z_{\rm{SN}}$ noticeably degrades the performance at $ z_{\rm{spec}} > 0.8$. 

In this section, we provided a variety of performance estimates for combinations of $z_{\mathrm{DCR}}$, $z_{\mathrm{Host}}$, and $z_{\mathrm{SN}}$. While it is difficult to precisely pinpoint how much improvement will result from combining $z_{\mathrm {DCR}}$ with photo-$z$'s for LSST as the survey strategy has yet to be finalized, our methodology of obtaining $z_{\mathrm{DCR}}$ from astrometry is robust. Future work will need to consider carefully how the $z_{\mathrm{DCR}}$ PDFs are combined with photo-$z$'s, which could involve machine learning algorithms that utilize insight gained from the use of OVL, FITPROB, as well as diagnostics shown in Appendix \ref{sec:Diagnostics_cosmology_prior}.

\section{Discussion and Conclusion} \label{sec:conclusion}

In this paper, we outlined a method to obtain astrometric redshifts of SNe Ia from multi-band, multi-epoch observations using Differential Chromatic Refraction or DCR, which is caused by the wavelength dependence of our atmosphere's refractive index. Because the DCR shifts depend on the source SED, it can in principle be used to infer its redshift if sufficient number of observations are available at moderately high air masses.

We evaluate our method on LSST-like simulations using an updated version of \verb|SNANA| incorporating statistical and systematic uncertainties and detailed calculations of the DCR effect.~We provide two types of point estimates, one where we take the 50th percentile of the cumulative distribution function of the posterior to be the estimated redshift (\CDF{}), and one where we take the peak of the posterior to be the estimated redshift (\Peaks{}).~We find that in the hypothetical cases of Perfect (no noise) simulations, $z_{\rm{DCR}}$ perform very well, although the estimate is not always accurate due to degenaracies. For the Realistic case, we find that our estimates at $z_{\rm{spec}} > 0.7$ are substantially degraded, but the $z_{\rm{DCR}}$ estimates are still quite reasonable at $z_{\rm{spec}} < 0.6$. 

We then combine $z_{\rm{DCR}}$ with host-galaxy photometric redshifts as well as SN Photometric redshifts using \verb|LCFIT+z|. As $z_{\rm{Host}}$ are completely independent from 
$z_{\rm{SN}}$ and $z_{\rm{DCR}}$, and $z_{\rm{DCR}}$ are close to independent from $z_{\rm{SN}}$, which we confirmed by calculating the Pearson correlation coefficients between them, combining the different methods results in substantial improvement in the redshift estimates.  Combining all three methods results in reduced bias, outlier fractions, and MAD deviation. The outlier fractions for $z_{\rm{SN}}$ is $\eta_{0.10} = 20.7\%$, while combining with both $z_{\rm{Host}}$ and reduces this to about 6.7\% for the all-redshifts sample.~Combining $z_{\rm{DCR}}$ on top of this further reduces $\eta_{0.10}$ to 6.0\%, and results in about 10\% more SN Ia candidates that pass selection cuts. 

We believe that our analysis is important in several ways.~First, this is the first demonstration of astrometric redshift measurements for SNe Ia. Contrary to \citet{kaczmarczik2009astro_z}, where the DCR shifts in the \textit{u} and \textit{g}-bands were treated as colors in the photo-$z$ algorithm, we constructed and exploited the full PDFs. Astrometric redshifts are especially useful since they are independent of host-galaxy photometric redshifts, and nearly independent from SN photometric redshifts, allowing the combination to result in more accurate redshift measurements.~$z_{\rm{DCR}}$ perform better at lower redshifts, complementing $z_{\rm{Host}}$, which perform better at higher redshifts. This will also allow us to use more SNe Ia for cosmology compared to using just SN + Host photometric redshifts, as well as reduce uncertainties.~While we are quite eager to see our new method used in a cosmology analysis, such an analysis is highly non-trivial and well beyond the scope of this work. 

Although we use the coadded DDF for LSST in this initial analysis, we expect that astrometric redshifts will also be useful for the Wide-Fast-Deep (WFD) survey where the chances of obtaining spectroscopic redshifts are much lower. While astrometric redshifts are not expected to provide much information at the higher redshifts, especially for the WFD due to low S/N and reduced cadence, we believe they will be similarly constraining as shown in this work for $z_{\rm{spec}} < 0.6$, which we have seen in preliminary analysis with the Dark Energy Survey (DES) shallow fields. Although we found in Section \ref{sec:realistic} that restricting to candidates with at least one observation with $\rm{AM} > 1.4$ does not improve $z_{\rm{DCR}}$ significantly for the DDF, a few intentional high AM observations might be beneficial for the Wide-Fast-Deep (WFD) fields as the number of observations per candidate are expected to be lower than the DDF on average.

We plan to implement astrometric redshifts of SNe Ia into our SNe Ia analysis pipeline for LSST observations, although we will have to account for the additional uncertainties and bias coming from the telescope instrumental properties discussed in Section \ref{sec:data_sims}. In this work, we did not consider atmospheric variability, although its effect on DCR shifts is likely to be small and it is possible to incorporate the recorded atmospheric conditions for each observation.~We recommend that LSST record the quantities required to calculate the atmospheric refraction index for each observation, such as air temperature, air pressure, and water vapor pressure.~Ideally, the index of refraction would be calculated for each observation and stored for later access. We also did not include uncertainties arising from host-galaxy modeling, which will make analysis in real observations more difficult.~Additionally, we note that the performance of astrometric redshifts will depend sensitively on the detailed survey strategy such as the cadence and the air mass distribution of the observations, which have not yet been finalized for LSST.~To prepare for using this new methodology in a future cosmology analysis on both simulations and real data, the DCR simulation tools have been integrated to \verb|SNANA|, and an improved scene-modeling photometry pipeline is underway.

Some additional potential applications of this work include: utilizing the PSF shape changes due to DCR and wavelength-dependent seeing to estimate redshifts as the 2D shape changes tend to be larger than the 1D DCR shifts, and extending astrometric redshifts to Type II supernovae. The former method is encouraging because the shape change caused by wavelength-dependent effects result in more apparent magnitude shifts than the positional shifts due to DCR, but the shape changes will almost certainly be more difficult to measure than 1D positional shifts.~Type II supernovae, like Type Ia supernovae, also display distinct emission lines which result in substantially different DCR shifts depending on the redshift. Thus, we expect that astrometric redshifts for Type II supernovae will also be useful as we have shown in this work for Type Ia supernovae. Another potential application could be to find mismatched host-galaxies using $z_{\rm{SN+DCR}}$. If the estimated $z_{\rm{SN+DCR}}$ is very different from $z_{\rm{Host}}$, the host-galaxy may not have not been identified correctly. This could be useful whether we are using host-galaxies for photometric redshifts or spectroscopic follow-up.

\begin{acknowledgments}

\noindent \textit{Acknowledgements.} This paper has undergone internal review in the LSST Dark Energy Science Collaboration.~We thank the internal reviewers Gautham Narayan and Robert Knop.~We also thank the anonymous referee for catching an error in the equation used to calculate the DCR shifts in the initial version of this draft, as well as helping the authors improve the clarity of this work.~LSST DESC acknowledges ongoing support from the Institut National de Physique Nucléaire et de Physique des Particules in France; the Science \& Technology Facilities Council in the United Kingdom; and the Department of
Energy, the National Science Foundation, and the LSST
Corporation in the United States. LSST DESC uses the
resources of the IN2P3/CNRS Computing Center (CCIN2P3–Lyon/Villeurbanne—France) funded by the Centre National de la Recherche Scientifique; the Univ. Savoie Mont Blanc—CNRS/IN2P3 MUST computing center; the National
Energy Research Scientific Computing Center, a DOE Office of Science User Facility supported by the Office of Science of the U.S. Department of Energy under contract No.~DE-AC02-05CH11231; STFC DiRAC HPC Facilities, funded by UK BIS
National E-infrastructure capital grants; and the UK particle
physics grid, supported by the GridPP Collaboration. This
work was performed in part under DOE contract DE-AC02-76SF00515.~J.L. and M.S. were supported by DOE grant DE-FOA-0002424 and NSF grant AST-2108094.~The work of A.I.M. was supported by Schmidt Sciences.

{\bf \noindent Author Contributions are Listed Below.}

J. Lee: led main analysis, wrote code and scripts for analysis, as well as the manuscript.~M. Sako: project initiator, supervision of lead graduate student, manuscript editing, administration, and management.~R. Kessler: updated {\sc SNANA} simulation to model atmospheric effects, 
light-curve fitting strategy, and manuscript editing.~A.I.~Malz:~generated the photometric redshift PDFs.

\end{acknowledgments}

%




\appendix \section{DCR calculations} \label{Appendix:DCR_calc}

\begin{figure*}
    \centering
    \includegraphics[width=0.98
\textwidth]{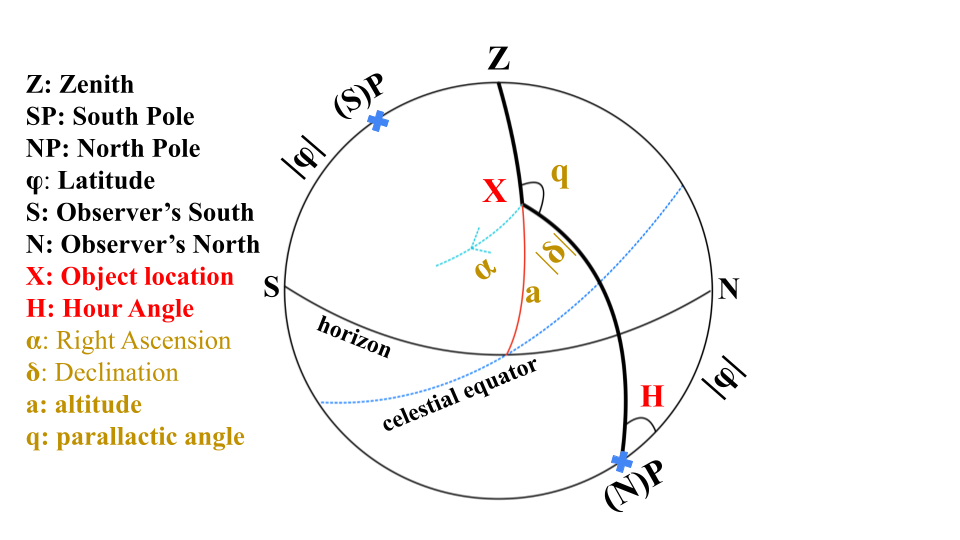}
    \caption{The celestial sphere for an observer in the Southern Hemisphere.}
    \label{fig:celestial_sphere}
\end{figure*}

In this section, we describe more explicitly how we implemented the DCR RA ($\alpha$) and DEC ($\delta$) shifts into \verb|SNANA| as mentioned in Section \ref{sec:data_sims}.~This is similar to the Appendix of \citet*{DES5YR-DCR} but for an observer in the Southern Hemisphere.~Figure \ref{fig:celestial_sphere} is a diagram of the celestial sphere for an observer (who would be in the center of the sphere) in the Southern Hemisphere, where Z is the zenith, (S)P and (N)P (denoted as just `P' from hereon) are the South and North celestial poles, and $\phi$ is the latitude of the observatory. Note that $\phi <0 $ for an observer in the Southern Hemisphere. This means that the angle between the South Pole and the observer's South or the angle between the North Pole and the observer's North is $|\phi|$. For an object located at X, $a$ is the altitude, $\alpha$ and $\delta$ are the RA and DEC, with $\delta < 0$ in this diagram, and $\alpha$ increases in the counterclockwise direction as viewed from above the \textbf{North} Pole. The angle XPZ or $H$, is the Hour Angle, and angle ZXP or $q$, is the parallactic angle, which are both typically defined using the North Celestial Pole.

In \verb|SNANA|, we calculate the parallactic angles using the spherical sine and cosine laws: 
\begin{equation} 
\begin{split}
    \frac{\sin(q)}{\sin(90^{\circ} + |\phi|)} = \frac{\sin(H)}{\sin(90^{\circ} - a)} = \frac{
    \sin(360^{\circ} - A)}{\sin(90^{\circ} - \delta)} \rightarrow \sin(q) =  \frac{\cos(\phi)\sin(H)}{\cos(a)} = \frac{-\cos(\phi)\sin(A)}{\cos(\delta)} \\ 
    \cos(90^{\circ} + |\phi|) = \cos(90^{\circ} - a) \cos(90^{\circ} + |\delta|) + \sin(90^{\circ} - a) \sin(90^{\circ} + |\delta|) \cos{q}, \rightarrow \cos(q) = \frac{\sin(\phi) - \sin(a)\sin(\delta)}{\cos(a)\cos(\delta)}
\end{split}
\end{equation} \label{eq:parallactic_angle}
\noindent where we have used that $|\phi| = -\phi$ and $|\delta| = -\delta$ and $A$ is the azimuth of the object. 

\begin{figure*}
    \centering
    \includegraphics[width=0.95
\textwidth]{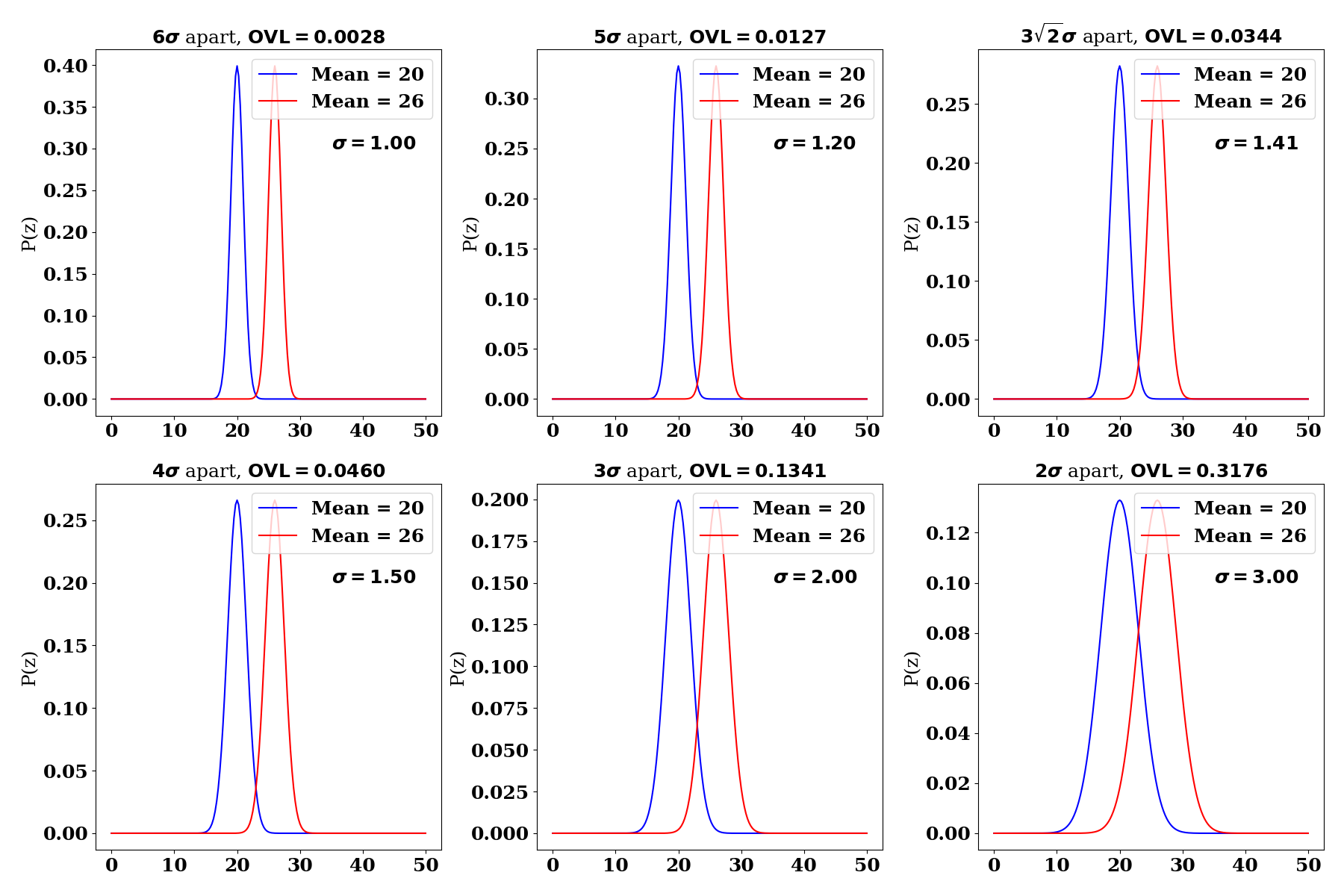}
    \caption{Overlapping Index depending on number of standard deviations apart for two identical normalized Gaussian distributions with different means. Because the means of the two distributions are fixed, the larger the $\sigma$, the higher the overlap. We take the case when the means of the two distributions are $3\sqrt{2}$ apart or when $\mathrm{OVL} > 0.0344$ as a threshold when combining $z_{\mathrm{DCR}}$ and $z_{\mathrm{Host}}$.}
    \label{fig:overlap_Gaussians}
\end{figure*}

Now, consider a DCR altitude shift towards the zenith, $\Delta a$. In flat Euclidean geometry, we would calculate the $\alpha$ and $\delta$ shifts as $\Delta \alpha = \Delta a \sin(180^{\circ} - q) = \Delta a \sin(q)$ and $|\Delta \delta| = \Delta a \cos(180^{\circ} - q) = - \Delta a \cos(q)$. However, these relations do not necessarily hold on a sphere. To calculate the $\alpha$ and $\delta$ shifts properly, we can utilize Napier's laws for right spherical triangles:
\begin{equation}
\begin{split}
    \sin(\Delta \alpha) = \sin(180^{\circ} - q) \sin(\Delta a) = \sin(q)\sin(\Delta a) \\ 
    \sin(|\Delta \delta|) = \sin(x) \sin(\Delta a) = \tan(\Delta \alpha) \cot(180^{\circ} - q)  = -\tan(\Delta \alpha)\frac{\cos (q)}{\sin(q)}
\end{split}\label{eq:DCR_shifts_appendix}
\end{equation} 

\noindent where the angle between $\Delta \alpha$ and $|\Delta \delta|$ is $90^{\circ}$, $|\Delta \delta|$ and $\Delta a$ is $180^{\circ}-q$, and $\Delta a$ and $\Delta \alpha$ is labeled as $x$. Since the shifts (altitude, RA, and DEC) we are considering are well below $1^{\circ}$, $\sin(\Delta \alpha) \approx \tan(\Delta \alpha) \approx \Delta \alpha$, and Equation \ref{eq:DCR_shifts_appendix} can be approximated as: 
\begin{equation}
    \Delta \alpha \approx \sin(q)\Delta a, \quad |\Delta \delta| \approx -\Delta \alpha \frac{\cos (q)}{\sin(q)} = - \cos(q) \Delta a
\end{equation}
\noindent Lastly, when we calculate the projected DCR shifts, we must account for the size of $1^{\circ}$ becoming smaller as we move farther away from the celestial equator, and $\Delta \delta$ is clearly negative for an upward DCR altitude shift, so we have $\Delta \alpha = \frac{\sin(q)}{\cos(\delta)} \Delta a$ and $\Delta \delta= \cos(q)\Delta a$ as in Equation \ref{eq:RA_DEC_DCR_shift}.

\section{In-depth Details from Methodology} \label{Appendix:Detailed_Methodology}

\subsection{Choice of reference star}

Using a combination of both \textit{Y}-band and \textit{z}-band average coordinates result in a slight, but noticeable improvement in terms of the outlier fractions (defined in Section \ref{sec:results}) particularly for the Perfect and systematics-only simulations from when using just \textit{z}-band average coordinates especially because we can utilize \textit{z}-band shifts when using the \textit{Y}-band average coordinates as the reference star coordinates. 

It is also possible to use the \textit{i}-band average coordinates as the reference star coordinates, which could be a better estimate than the \textit{z} or \textit{Y}-band average coordinates since the S/N tends to be much higher in the \textit{i}-band. However, we also need to consider the fact that DCR shifts tend to be larger in the \textit{i}-band compared to the \textit{zY}-bands, sometimes to a non-negligible degree. If a high S/N \textit{i}-band observation happens to have a large DCR shift, the average \textit{i}-band coordinates for that particular candidate will no longer be a good proxy for the reference star coordinates.~Furthermore, using the \textit{i}-band average coordinates could bias the \textit{iz}-band DCR shift measurements and using only the \textit{ugr}-band DCR shifts reduces constraining power.~We leave to future works to investigate in more detail which is most appropriate to use as the reference star coordinates.

\subsection{$\chi^2$ denominator for Perfect simulations}

When computing the posterior PDFs for Perfect ($\sigma_{\rm{syst}} = \sigma_{\rm{stat}} = 0$) simulations with or without marginalization of $c$ or $x_1$ (Section \ref{sec:sigma_syst only}), the denominator of Equation (9) 
is zero in theory, and $\chi^2$ becomes infinite while $\mathcal{P}(z) = 0$. To avoid this numerical artifact, we calculate $\mathcal{P}(z)$ with $\sigma_{\rm{tot}} = 1.0$ mas. \Peaks{} estimates are independent of the $\sigma_{\rm{tot}}$ value.~We note that the \CDF{} estimates are sensitive to our choice of $\sigma_{\rm{tot}}$, with a smaller $\sigma_{\rm{tot}}$ resulting in the \CDF{} value becoming closer to the \Peaks{} value. This is because an overall multiplicative factor ($ > 1$) in the $\chi^2$, corresponding to smaller $\sigma_{\rm{tot}}$, translates to the highest amplitude peak in $\mathcal{P}(z)$ becoming more distinct. We confirmed that as $\sigma_{\rm{tot}}$ decreases, the \CDF{} estimates increasingly match that of \Peaks{}. Additionally, the uncertainties converge to zero as $\sigma_{\rm{tot}}$ decreases. Therefore, we show the same (\Peaks{}) results for \CDF{} and \Peaks{} for the Perfect simulations without uncertainties where $\sigma_{\rm{tot}}$ should be as close to zero as possible.~While taking $\sigma_{\rm{tot}} = 1.0$ mas renders $\mathcal{P}(z) = 0$ for all redshifts for a small fraction of candidates when marginalization is included, the percentages of this occurring is at most 0.5\% of the candidates that pass the epoch cut. 

\section{Overlapping Index} \label{Appendix:Overlap}

In Section \ref{sec:methods}, we defined the Overlapping Index (OVL) as:
\begin{equation}
    \mathrm{OVL} (p_0, p_1) = \int \min(p_0(x), p_1(x)) dx
\end{equation}

\noindent where $p_0(x)$ and $p_1(x)$ are PDFs. In Figure \ref{fig:overlap_Gaussians}, we show two identical normalized Gaussian distributions with means of 20 (red) and 26 (blue), depending on their standard deviations ($\sigma$) and hence number of $\sigma$'s the means of the two distributions are apart. When normalized as shown here, we can say that the blue curve overlaps 3.44\% with the red curve when their means are $3\sqrt{2} \sigma$ apart. In the main text, we take $\mathrm{OVL} > 0.0344$ to be the threshold for two PDFs being compatible with each other.~As the means of the two distributions become farther away from each other in terms of $\sigma$, the Overlapping Index decreases, and we see smaller overlap.~Note that the PDFs shown in Figure~\ref{fig:overlap_Gaussians} are not representative of our $z_{\rm DCR}$ or $z_{\rm Host}$ PDFs; they are meant to illustrate the utility of OVL in a simple but intuitive way.

~While it is possible to compute the overlapping index assuming that $z_{\rm{SN}}$ PDFs are Gaussian (although this is not true as shown in \citet{sako2018sdssII}), we found that the OVL values between $z_{\rm{SN}}$ and $z_{\rm{DCR}}$ or $z_{\rm{Host}}$ are very small, due to the $z_{\rm{SN}}$ $1\sigma$ error bars from \verb|LCFIT+z| being small, with a mean of $0.017$ and median of $0.011$. The \verb|LCFIT+z| $z_{\rm{SN}}$ error bars are computed using the default {\sc MINUIT} \citep{james1975minuit} error option (MIGRAD, `Migration by Gradientes').~Better photo-$z$ error estimates can be obtained, such as by using the MINOS (`Minimization by Orthogonal Simplex') option in MINUIT, or MCMC fitting, but the fitting speed would be significantly slower.

\section{Cosmology Prior for Light-Curve Fitting} \label{Appendix:cosmology_prior}

In Section \ref{sec:data_sims}, we use a weak cosmology prior for \verb|LCFIT+z| to obtain $T_{\rm{obs}}$ for Realistic $z_{\rm{DCR}}$. Additionally, we use a cosmology prior in Section \ref{sec:Diagnostics_cosmology_prior} to see improvement in the $z_{\rm{SN}}$ performance. Here, following \citet{kessler2010photo_z} and \citet{moller2024}, we describe how the prior is applied. 

We begin with the approximate fitted distance modulus, $\mu_{\rm{SALT3}}$:
\begin{equation}
    \mu_{\rm{SALT3}} = 30.0 - 2.5 \log_{10}{x_0} + \alpha x_1 - \beta c
\end{equation}

\noindent with $x_0$ (amplitude), $x_1$ (stretch), and $c$ (color) being SALT3 light-curve parameters. We use $\alpha = 0.14$ and $\beta = 3.2$ as our default. The difference between the fitted and theoretical distance modulus, $\mu_{\rm{DIFF}}$, is given by:

\begin{equation}
    \mu_{\rm{DIFF}} = \mu_{\rm{SALT3}} - \mu_{\rm{th}}(z_{\rm{SN}})
\end{equation}

\noindent where $\mu_{\rm{th}}$ is the theoretical distance modulus. The distance modulus uncertainty, $\sigma_{\mu}$, can be computed using:
\begin{equation}
    \sigma_{\mu}^2 = \bigg(\frac{d\mu}{d\Omega_{m}} \Omega_{m,\rm{ERR}}\bigg)^2 + \bigg(\frac{d\mu}{dw} w_{\rm{ERR}}\bigg)^2
\end{equation}

\noindent where $\Omega_{m,\rm{ERR}}$ and $w_{\rm{ERR}}$ are uncertainties for $\Omega_{m}$ and $w$ respectively. As mentioned in Section \ref{sec:Diagnostics_cosmology_prior}, we use $\Omega_{m,\rm{ERR}} = 0.03$ and $w_{\rm{ERR}} = 0.1$. The reduced $\chi_{\rm{SALT3}}^2$ is then given by: 

\begin{equation}
    \chi_{\rm{SALT3}}^2 = \bigg( \frac{\mu_{\rm{DIFF}}}{\sigma_{\mu}} P_{\mu\rm{ERRSCALE}}  \bigg)^2.
\end{equation}

\noindent $P_{\mu\rm{ERRSCALE}}$ is the distance modulus prior: a Gaussian profile with a standard deviation of $2\sigma_{\mu}$ when obtaining $T_{\rm{obs}}$ in Section \ref{sec:data_sims}, and $1\sigma_{\mu}$ when assessing improvement in the $z_{\rm{SN}}$ performance in Section \ref{sec:Diagnostics_cosmology_prior}.

\begin{table*}[]
    \centering
\begin{tabular}{ccccccc}
\toprule
                    Type &           Bias  (CDF50) &          Bias (PDFPeak) & $\eta_{0.10}$ (CDF50) & $\eta_{0.10}$ (PDFPeak) & $\sigma_{\rm{MAD}}$ (CDF50) & $\sigma_{\rm{MAD}}$ (PDFPeak) \\
\midrule
        No selection cut & -0.103$\pm$0.001 & -0.067$\pm$0.001 &               0.577 &                 0.444 &                     0.139 &                       0.123 \\
                 Default & -0.048$\pm$0.001 & -0.024$\pm$0.002 &               0.333 &                 0.335 &                     0.086 &                       0.077 \\
        Default + AM cut & -0.041$\pm$0.001 & -0.020$\pm$0.002 &               0.306 &                 0.313 &                     0.080 &                       0.069 \\
              Fmax-clump & -0.048$\pm$0.001 & -0.025$\pm$0.002 &               0.344 &                 0.354 &                     0.090 &                       0.085 \\
Ideal $T_{\mathrm{obs}}$ & -0.050$\pm$0.001 & -0.027$\pm$0.002 &               0.320 &                 0.324 &                     0.083 &                       0.072 \\
\bottomrule
\end{tabular}
    \caption{Bias and error on the mean, $\eta_{0.10}$, and $\sigma_{\rm{MAD}}$ for the analysis methods indicated in the first column. With stricter selection cuts, we see improved metrics. Using peak MJDs from LCFIT+z (which is mostly the case for `Default') improves performance compared to using `Fmax-clump.'}
    \label{tab:statistics_realistic_by_selection_cut}
\end{table*}

\section{Additional Results}

\subsection{Sensitivity of Astrometric Redshifts on Selection Cuts}\label{sec:realistic_cuts}

We show in Table \ref{tab:statistics_realistic_by_selection_cut} the bias, $\eta_{0.10}$, and $\sigma_{\rm{MAD}}$ depending on the selection criteria: no selection cuts, default selection cut (42.9\% of total candidates) and default + AM selection cut (37.6\% of total candidates) where the AM cut requires at least one observation at $\rm{AM} > 1.4$ for a given candidate.~As expected, the performance of $z_{\rm{DCR}}$ improves with more stringent cuts. We also see larger bias in \CDF{} (compared to \Peaks{}) for all three types of cuts, which is expected given that \CDF{} estimates most $z_{\rm{spec}} > 0.7$ SNe to be at $z_{\rm{DCR}} \approx 0.6$. The outlier fractions are similar for \CDF{} and \Peaks{}, with the exception of `No selection cut,' where the bias for \CDF{} is significantly higher, due to the inclusion of disproportionately more high-$z$ SNe Ia.~Additionally requiring the AM cut improves the estimates, but not significantly. While \citet{kaczmarczik2009astro_z} suggests that a number of moderately high AM observations will result in considerably more accurate redshift estimates for quasars, we note that for the LSST DDF SNe Ia simulations, most of the candidates that pass the default selection cut also pass the AM cut as there are multiple (usually over 5) observations per SN.~We also show the metrics depending on the choice of $T_{\rm{obs}}$, and as expected, using peak MJDs from \verb|LCFIT+z| results in some improvement from the `Fmax-clump' method, but less than the Ideal $T_{\rm{obs}}$ case, where we use the exact peak MJDs.
\subsection{Low-$z$ vs. High-$z$ Improvements when $z_{\rm{SN}}$ is combined}\label{sec:zSN+DCR+Host_low-z_high-z}

In Table \ref{tab:statistics_low_high_z} (where we show the bias, $\eta_{0.10}$ and $\sigma_{\rm{MAD}}$ at low-$z$ and high-$z$ separately for \Peaks{} when not combining with $z_{\rm{SN}}$), both the bias and outlier fractions are much lower at low-$z$ than high-$z$ for $z_{\rm{DCR}}$, while the outlier fractions are much lower at high-$z$ for $z_{\rm{Host}}$ as seen earlier. $z_{\rm{SN}}$ show larger bias at low-$z$ but a much smaller outlier fraction (10.4\%) than at high-$z$ (26.3\%) as with $z_{\rm{DCR}}$.~Combining $z_{\rm{SN}}$ with $z_{\rm{Host}}$ reduces both the bias and outlier fractions significantly at both low-$z$ and high-$z$.~While it is no surprise that the outlier fraction at high-$z$ is about a third of that for $z_{\rm{SN}}$ (8.3\%), it is intriguing that the outlier fraction at low-$z$ is also 7\% lower, at 3.3\% given that the $z_{\rm{Host}}$-only outlier fraction at low-$z$ is 42.5\%.~Combining $z_{\rm{SN}}$ with $z_{\rm{DCR}}$ greatly reduces the bias and outlier fraction at low-$z$, with the outlier fraction being 2.3\%, but the bias at high-$z$ is only a little smaller than $z_{\rm{DCR}}$ at high-$z$, and the outlier fraction at high-$z$ is reduced by a small amount (22.7\%) as one would expect.~When all three methods are combined, the bias at both low-$z$ and high-$z$ are close to zero, and the outlier fractions are similarly low at 1.9\% and 8.5\% for low-$z$ and high-$z$ respectively, although using $z_{\rm{SN+DCR}}$ at low-$z$ and $z_{\rm{SN+Host}}$ at high-$z$ shows similar performance.~A similar trend can be seen with $\sigma_{\rm{MAD}}$ as well; $z_{\rm{DCR}}$ has a much lower value at low-$z$ while $z_{\rm{Host}}$ has a lower value at high-$z$, and hence combining each with $z_{\rm{SN}}$ displays lower (or at least similar) values at low-$z$ and high-$z$ respectively. $z_{\rm{SN+DCR+Host}}$ has the lowest $\sigma_{\rm{MAD}}$ values for both low-$z$ and high-$z$. 

\begin{table*}[]
    \centering
\begin{tabular}{ccccccc}
\toprule
                  Type &           Bias (low-z) &           Bias (high-z) & $\eta_{0.10}$ (low-z) & $\eta_{0.10}$ (high-z) & $\sigma_{\rm{MAD}}$ (low-z) & $\sigma_{\rm{MAD}}$ (high-z) \\
\midrule
        $z_{\rm{DCR}}$ & 0.007$\pm$0.002 & -0.043$\pm$0.002 &                 0.162 &                  0.437 &                       0.033 &                        0.123 \\
       $z_{\rm{Host}}$ & 0.006$\pm$0.003 & -0.054$\pm$0.002 &                 0.425 &                  0.268 &                       0.123 &                        0.065 \\
   $z_{\rm{DCR+Host}}$ & 0.008$\pm$0.001 & -0.025$\pm$0.002 &                 0.163 &                  0.203 &                       0.041 &                        0.055 \\
         $z_{\rm{SN}}$ & 0.019$\pm$0.002 &  0.001$\pm$0.001 &                 0.104 &                  0.263 &                       0.008 &                        0.059 \\
    $z_{\rm{SN+Host}}$ & 0.006$\pm$0.001 & -0.004$\pm$0.001 &                 0.033 &                  0.086 &                       0.008 &                        0.023 \\
     $z_{\rm{SN+DCR}}$ & 0.003$\pm$0.001 & -0.034$\pm$0.001 &                 0.023 &                  0.227 &                       0.007 &                        0.041 \\
$z_{\rm{SN+DCR+Host}}$ & 0.003$\pm$0.001 & -0.007$\pm$0.001 &                 0.019 &                  0.085 &                       0.007 &                        0.021 \\
\bottomrule
\end{tabular}
    \caption{Bias and error on the mean, $\eta_{0.10}$, and $\sigma_{\rm{MAD}}$ for low-$z$ ($z_{\rm{spec}} < 0.6$) and high-$z$ ($z_{\rm{spec}} > 0.6$) for various combinations of the three methods discussed in this work. For $z_{\rm{DCR}}$, $z_{\rm{Host}}$, and $z_{\rm{DCR+Host}}$, we show the \Peaks{} values.
 $z_{\rm{DCR}}$ shows good performance at low-$z$ while $z_{\rm{Host}}$ shows good performance at high-$z$, which leads to $z_{\rm{SN+DCR}}$ being significantly improved at low-$z$ compared to $z_{\rm{SN}}$ and $z_{\rm{SN+Host}}$ being significantly improved at high-$z$. $z_{\rm{SN+DCR+Host}}$ show similar performance to $z_{\rm{SN+DCR}}$ and $z_{\rm{SN+Host}}$ at low-$z$ and high-$z$ respectively.}
    \label{tab:statistics_low_high_z}
\end{table*}

\section{Diagnostics and Cosmology Prior} \label{sec:Diagnostics_cosmology_prior}

As discussed at the end of Section \ref{sec:data_sims}, we impose a $\mathrm{FITPROB} \ge 0.01$ selection cut when combining with $z_{\mathrm{SN}}$, which addresses some of the compatibility issues between the different redshift estimation methods. While this is sufficient for our initial analysis with simulations, we dive deeper and present some diagnostics as well as how much improvement can result from using a cosmology prior for \verb|LCFIT+z|. 

\begin{figure*}
    \centering
    \includegraphics[width=0.99\textwidth]{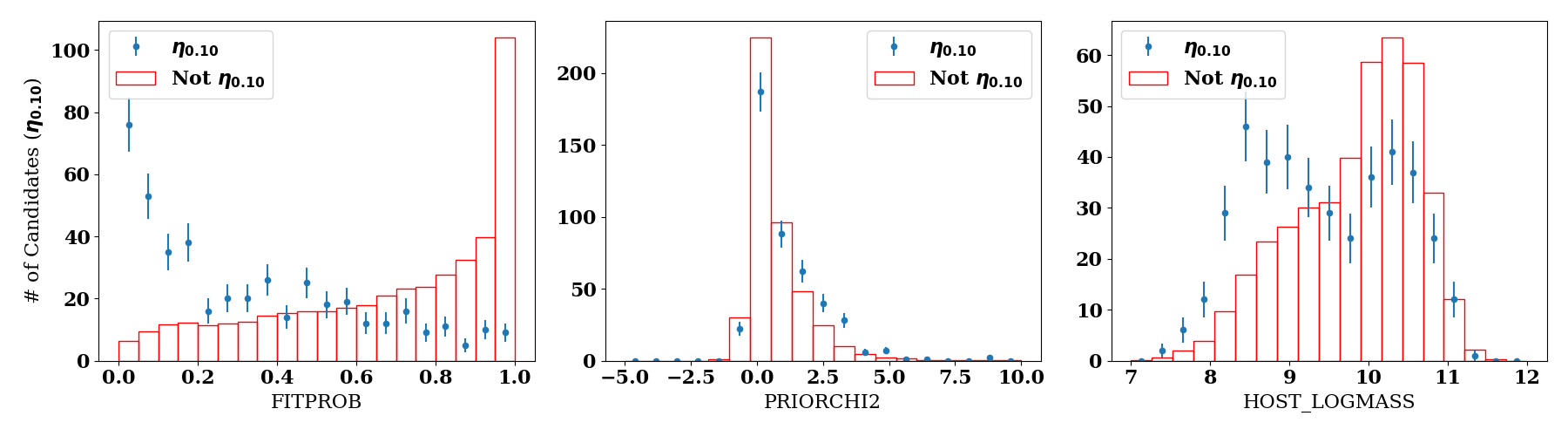}
    \caption{Diagnostics for $z_{\mathrm{SN+DCR+Host}}$. The blue points show the distributions for outliers, while the red histograms show the \textit{relative} number of candidates for non-outliers. The red bars have been scaled such that the area under the $\eta_{0.10}$ histograms and Not $\eta_{0.10}$ histograms are equal. We show the error bars for the $\eta_{0.10}$ points only because Not $\eta_{0.10}$ has many more candidates than $\eta_{0.10}$ and hence smaller error bars. As expected, outliers have lower FITPROB than non-outliers as well as slightly higher prior-$\chi^2$. Additionally, outliers are more likely to occur in lower mass host galaxies.}
    \label{fig:Diagnostics}
\end{figure*}

In Figure \ref{fig:Diagnostics}, we show some diagnostics histograms for $z_{\mathrm{SN+DCR+Host}}$ (with the $\mathrm{FITPROB} \ge 0.01$ cut and the compatibility cut for the $z_{\mathrm{DCR+Host}}$ prior) for candidates that are within $\eta_{0.10}$ and outside $\eta_{0.10}$: FITPROB, prior-$\chi^2$, and host-galaxy $\log$ mass. For each of the panels, the blue points show the number of candidates within the $x$-axis bins for outliers, while the red histogram shows the distribution of non-outlier (Not $\eta_{0.10}$) candidates, scaled to have the same integral as the outlier distribution. 

On the left panel, we see that $\eta_{0.10}$ has more candidates at lower FITPROB while Not $\eta_{0.10}$ display the opposite, as expected. Adding useful information from FITPROB into a more complex machine learning algorithm could therefore improve the selection of events. In the middle, we see that the prior-$\chi^2$ histograms are similar for $\eta_{0.10}$ and Not $\eta_{0.10}$, although $\eta_{0.10}$ is skewed towards slightly higher prior-$\chi^2$ values.~On the right panel, $\eta_{0.10}$ is shown to have more low host mass contribution compared to Not $\eta_{0.10}$. This could be due to low-mass host galaxies having lower S/N at high redshifts, degrading all three redshift estimates. 


\begin{figure*}
      \centering
  \begin{minipage}[b]{0.45\textwidth}
    \includegraphics[width=\textwidth]{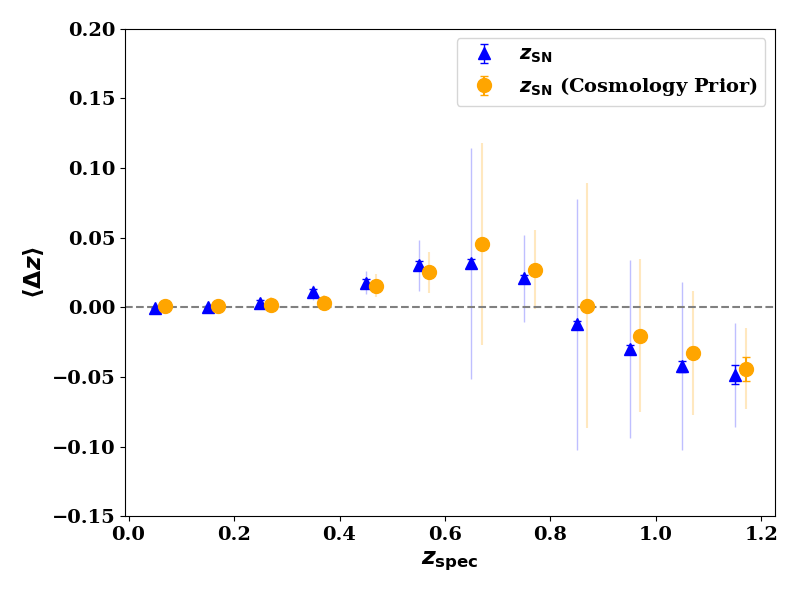}
    \caption{Binned residuals (bias), error on the mean (bold error bars) and the size of $\sigma_{\mathrm{MAD}}$ (light error bars) as a function of $z_{\rm{spec}}$ for $z_{\rm{SN}}$, with and without cosmology priors. Using a cosmology prior generally tends to reduce the bias and $\sigma_{\mathrm{MAD}}$ except at $0.6 < z_{\rm{spec}} < 0.8$.}
    \label{fig:mean_residuals_z_SN_cosmology_prior}
  \end{minipage}
  \hfill
  \begin{minipage}[b]{0.45\textwidth}
    \includegraphics[width=\textwidth]{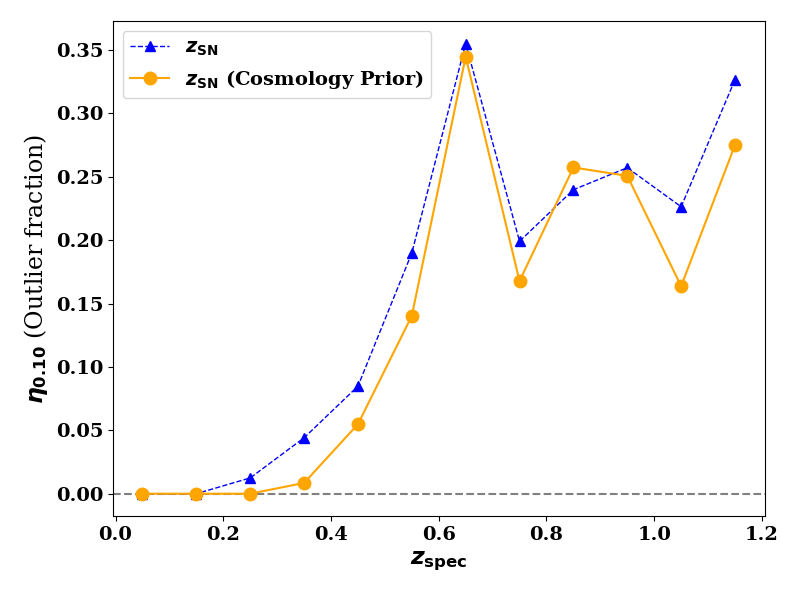}
    \caption{$\eta_{0.10}$ as a function of $z_{\rm{spec}}$ for $z_{\rm{SN}}$, with and without cosmology priors. Similar to the bias and $\sigma_{\mathrm{MAD}}$, a cosmology prior generally reduces $\eta_{0.10}$.}
    \label{fig:outlier_rates_z_SN_cosmology_prior}
  \end{minipage}
\end{figure*}



For all of the aforementioned results, we did not impose any cosmology prior apart from determining $T_{\mathrm{obs}}$ using \verb|LCFIT+z| for the Realistic case (Section \ref{sec:data_sims}). Here, we investigate potential improvement from using a weak cosmology prior for \verb|LCFIT+z|.~More detail on how the prior is applied is given in Appendix \ref{Appendix:cosmology_prior}. In Figures \ref{fig:z_pred_z_SN_cosmology_prior} - \ref{fig:outlier_rates_z_SN_cosmology_prior}, we show the performance of $z_{\rm{SN}}$ only with a weak cosmology prior of $w = -1.0 \pm 0.1$ and $\Omega_{m} = 0.3 \pm 0.03$. In Figure \ref{fig:z_pred_z_SN_cosmology_prior}, we see that compared to the top panel of Figure \ref{fig:SN_photo-z_combined}, the outlier fractions have noticeably decreased, $\eta_{0.10}$ is 0.205 and 0.186 without and with the cosmology prior respectively for all redshifts, while for $z < 0.6$ only, $\eta_{0.10}$ is 0.094 and 0.052 respectively. Note that the number of candidates that pass the selection cuts (\verb|LCFIT+z| convergence + epoch and S/N cut + $\mathrm{FITPROB} \ge 0.01$ cut) have also increased by 7\% compared to the case without the cosmology prior. In Figures \ref{fig:mean_residuals_z_SN_cosmology_prior} and \ref{fig:outlier_rates_z_SN_cosmology_prior}, we show that the bias, $\eta_{0.10}$, as well as $\sigma_{\mathrm{MAD}}$ are generally slightly reduced with the exception of some bins between $0.6 < z_{\rm{spec}} < 0.9$.~This additional analysis shows that in practice, we can utilize a weak cosmology prior to obtain better estimates for both photometric and astrometric redshifts. 

\begin{figure*}
    \centering
    \includegraphics[width=0.96\textwidth]{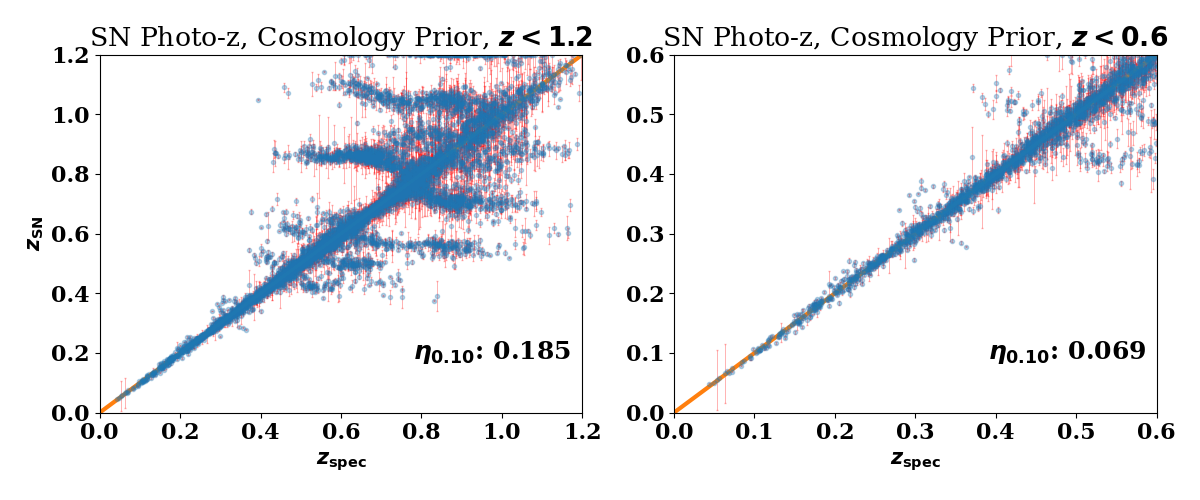}
    \caption{$z_{\rm{SN}}$ vs. $z_{\rm{spec}}$ with a weak cosmology prior for LCFIT+z. Compared to the top panels of Figure \ref{fig:SN_photo-z_combined} where there is no cosmology prior, the estimates are closer to the true values throughout all redshifts and the outlier fractions are lower.}
    \label{fig:z_pred_z_SN_cosmology_prior}
\end{figure*}

\section{Systematic Effects} \label{sec:sigma_syst only}



\begin{table*}[]
    \centering
\begin{tabular}{ccccccc}
\toprule
                         Type &            Bias (CDF50) &          Bias (PDFPeak) & $\eta_{0.10}$ (CDF50) & $\eta_{0.10}$ (PDFPeak) & $\sigma_{\rm{MAD}}$ (CDF50) & $\sigma_{\rm{MAD}}$ (PDFPeak) \\
\midrule
                      Perfect & -0.003$\pm$0.000 & -0.003$\pm$0.000 &               0.014 &                 0.014 &                     0.008 &                       0.008 \\
 $\sigma_{\rm{syst}} = 2$ mas & -0.007$\pm$0.000 & -0.003$\pm$0.000 &               0.015 &                 0.016 &                     0.009 &                       0.009 \\
 $\sigma_{\rm{syst}} = 5$ mas & -0.010$\pm$0.000 & -0.004$\pm$0.000 &               0.026 &                 0.023 &                     0.011 &                       0.011 \\
$\sigma_{\rm{syst}} = 10$ mas & -0.019$\pm$0.000 & -0.006$\pm$0.000 &               0.069 &                 0.044 &                     0.019 &                       0.016 \\
             $c$ marginalized & -0.007$\pm$0.000 & -0.004$\pm$0.000 &               0.011 &                 0.016 &                     0.006 &                       0.007 \\
            $x_1$ marginalized & -0.006$\pm$0.000 & -0.003$\pm$0.000 &               0.012 &                 0.014 &                     0.007 &                       0.008 \\
     realistic $T_{\rm{obs}}$ & -0.005$\pm$0.000 & -0.002$\pm$0.000 &               0.040 &                 0.040 &                     0.010 &                       0.010 \\
       All syst for realistic & -0.010$\pm$0.001 & -0.004$\pm$0.001 &               0.062 &                 0.069 &                     0.015 &                       0.014 \\
\bottomrule
\end{tabular}
    \caption{Bias, outlier fractions, and $\sigma_{\rm{MAD}}$ for various systematics. The impact of $\sigma_{\rm{syst}} = 2$ or $5$ mas, marginalization over $c$ and $x_1$ is small, while using realistic $T_{\rm{obs}}$ results in more degradation. Larger $\sigma_{\rm{syst}}$ results in substantially larger degradation compared to other systematics. `All syst. for realistic' means using realistic $T_{\rm{obs}}$, marginalizing over $c$ and $x_1$, and assuming $\sigma_{\rm{syst}} = 5$ mas. }
    \label{tab:statistics_syst_only}
\end{table*}

Since we have assessed the Perfect and Realistic cases in Sections \ref{sec:Perfect} and \ref{sec:realistic}, here we show how sensitive our $z_{\rm{DCR}}$ results are to various levels of systematic effects in the absence of statistical uncertainties.~The systematic uncertainties considered here include \textit{unmodeled} telescopic effects as discussed in Section \ref{sec:data_sims}, and we assume that systematics leading to biases can typically be modeled and accounted for in a real survey.~In Table \ref{tab:statistics_syst_only}, we show again the bias, $\eta_{0.10}$, and $\sigma_{\rm{MAD}}$ for the Perfect case as we saw in Section \ref{sec:Perfect} as reference. We also show the metrics for a Perfect simulation with the following systematic variations: $\sigma_{\rm{syst}} = 2$ mas,  $\sigma_{\rm{syst}} = 5$ mas, $\sigma_{\rm{syst}} = 10$ mas, $c$ marginalized, $x_1$ marginalized, and realistic $T_{\rm{obs}}$. For each systematic we tested, all the other conditions were taken to be the same as the Perfect case.~As described in Section \ref{sec:methods}, \CDF{} and \Peaks{} results are the same when $\sigma_{\rm{tot}}$ should be zero; so for Perfect, $c$ marginalized, $x_1$ marginalized, and realistic $T_{\mathrm{obs}}$.

First, we show that $\sigma_{\rm{syst}} = 5$ mas, which we took to be our default systematic floor throughout Section \ref{sec:realistic}, only causes a slight degradation compared to the Perfect case with $\eta_{0.10}$ being around 2.6\% and $\sigma_{\rm{MAD}}$ being around 0.011 for both \CDF{} and \Peaks{}.~\CDF{} and \Peaks{} show similar performance in terms of $\eta_{0.10}$ and $\sigma_{\rm{MAD}}$, which indicates that the degradation primarily occurs at the highest redshifts where the PDFs tend to be flatter than at lower redshifts, meaning that systematic effects can smear them out enough to result in incorrect redshift estimates. Bias values are farther away from zero for \CDF{} than \Peaks{}, closely resembling the Realistic case. 

Marginalizing over $x_1$ and $c$, as well as taking $\sigma_{\rm{syst}} = 2$ mas does not change the performance metrics much, suggesting that the level of degeneracies that result in incorrect redshift estimates at the high redshifts as shown in Figure~\ref{fig:LSST_z_pred_vs_z_true_ideal} remain similar.


Using realistic $T_{\rm{obs}}$ (mostly using \verb|LCFIT+z|, with 11,595 candidates remaining after \verb|LCFIT+z| out of the 13,827 that pass the Epoch cut) causes a larger degradation, with $\eta_{0.10}$ at about 4.0\%, or around two to three times that of the Perfect case while $\sigma_{\rm{MAD}}$ also increases but to a smaller extent. The bias values are similar to the Perfect case, but have larger error bars.~We find that the $z_{\rm{DCR}}$ performance with all of the systematics included in the Realistic case (using realistic $T_{\rm{obs}}$, marginalizing over $c$ and $x_1$, and assuming $\sigma_{\rm{syst}} = 5$ mas), but with $\sigma_{\rm{syst}} = 0$, is considerably better than the Realistic case with $\eta_{0.10} = 6.9\%$ and $\sigma_{\rm{MAD}} = 0.014$ for \Peaks{} (similar values for \CDF{}) and the bias being $-0.010\pm0.001$ and $-0.004\pm0.001$ for \CDF{} and \Peaks{} respectively, meaning that our $z_{\rm{DCR}}$ estimates are mostly limited by S/N. 

As mentioned in Section \ref{sec:data_sims}, we note that our assumption of a $\sigma_{\rm{syst}} = 5$ mas floor is reasonable and possible to achieve for LSST given the DECam value of 3--6 mas \citep{bernstein2017astrometric}. Additionally, we consider the $z_{\rm{DCR}}$ performance when $\sigma_{\rm{syst}}$ is 2 mas and 10 mas. In Table \ref{tab:statistics_syst_only}, we show that $\sigma_{\rm{syst}} = 2$ mas results in $\eta_{0.10}$ being about half that of $\sigma_{\rm{syst}} = 5$ mas or similar to the Perfect case while the bias and $\sigma_{\rm{MAD}}$ similarly improve slightly. $\sigma_{\rm{syst}} = 10$ mas results in $\eta_{0.10}$ being about 3 to 4 times larger than the Perfect case at 6.9\% and 4.4\% for \CDF{} and \Peaks{} and the bias and $\sigma_{\rm{MAD}}$ also degrade substantially more compared to the other systematics considered here. The larger discrepancy in the metrics between \CDF{} and \Peaks{} for $\sigma_{\rm{syst}} = 10$ mas suggests that the high-$z$ estimates are affected disproportionately. The metrics we present here for larger $\sigma_{\rm{syst}}$ shows that limiting $\sigma_{\rm{syst}}$ to 5 mas or lower will be crucial when implementing $z_{\rm{DCR}}$ for LSST. 




\bibliography{refs}{}
\bibliographystyle{yahapj_twoauthor_amp} 




\end{document}